\documentclass[review, 5p, times, authoryear]{elsarticle}

\usepackage{amsthm}
\usepackage{amsmath}
\usepackage{esint}
\usepackage{cancel}
\usepackage{comment}
\usepackage{threeparttable}
\usepackage{multirow}
\usepackage{rotating}
\usepackage{adjustbox}
\usepackage{multicol}
\renewcommand{\vec}[1]{\mathbf {#1}}
\renewcommand{\d}{\textnormal{d}}
\journal{J. Wind. Eng. Ind. Aerodyn.}

\usepackage{url}
\usepackage{framed} 
\usepackage{nomencl} 
\usepackage{longtable}
\usepackage{graphicx}

\makenomenclature

\renewcommand*\nompreamble{\begin{multicols}{2}}

\renewcommand*\nompostamble{\end{multicols}}

\begin{document}
\begin{frontmatter}

\title{The drag length is key to quantifying tree canopy drag}

\author[icl]{Dipanjan Majumdar\corref{cor1}}
\author[ramboll]{Giulio Vita}
\author[arup]{Rubina Ramponi}
\author[wsp]{Nina Glover}
\author[icl]{Maarten van Reeuwijk}

\cortext[cor1]{d.majumdar@imperial.ac.uk}

\affiliation[icl]{organization={Department of Civil and Environmental Engineering, Imperial College London},
            addressline={South Kensington Campus}, 
            city={London},
            postcode={SW7 2AZ}, 
            country={United Kingdom}}

\affiliation[ramboll]{organization={RTR Advanced Simulations UK, Ramboll UK Ltd.},
            addressline={Cornerblock Two, 2 Cornwall St}, 
            city={Birmingham},
            postcode={B3 2DX}, 
            country={United Kingdom}}

\affiliation[arup]{organization={Specialist Technology, Analytics and Research, Arup},
            addressline={50 Ringsend Road}, 
            city={Dublin},
            postcode={D04 T6X0}, 
            country={Ireland}}

\affiliation[wsp]{organization={WSP},
            addressline={3 Wellington Place}, 
            city={Leeds},
            postcode={LS1 4AP}, 
            country={United Kingdom}}

\begin{abstract}

The effects of trees on urban flows are often determined using computational fluid dynamics approaches which typically use a quadratic drag formulation based on the leaf-area density $a$ and a volumetric drag coefficient $C_{d}^V$ to model vegetation. In this paper, we develop an analytical model for the flow within a vegetation canopy and identify that the drag length $\ell_d = (a C_d^V)^{-1}$ is the key metric to describe the local tree drag characteristics. A detailed study of the literature suggests that the median $\ell_d$ observed in field experiments is $21$ m for trees and $0.7$ m for low vegetation (crops). A total of $168$ large-eddy simulations are conducted to obtain a closed form of the analytical model. The model allows determining $a$ and $C_d^V$ from wind-tunnel experiments that typically present the drag characteristics in terms of the classical drag coefficient $C_d$ and the aerodynamic porosity $\alpha_L$. We show that geometric scaling of $\ell_d$ is the appropriate scaling of trees in wind tunnels. Evaluation of $\ell_d$ for numerical simulations and wind-tunnel experiments (assuming geometric scaling $1:100$) in literature shows that the median $\ell_d$ in both these cases is about $5$ m, suggesting possible overestimation of vegetative drag.

\end{abstract}

\begin{graphicalabstract}
\includegraphics[scale=0.2]{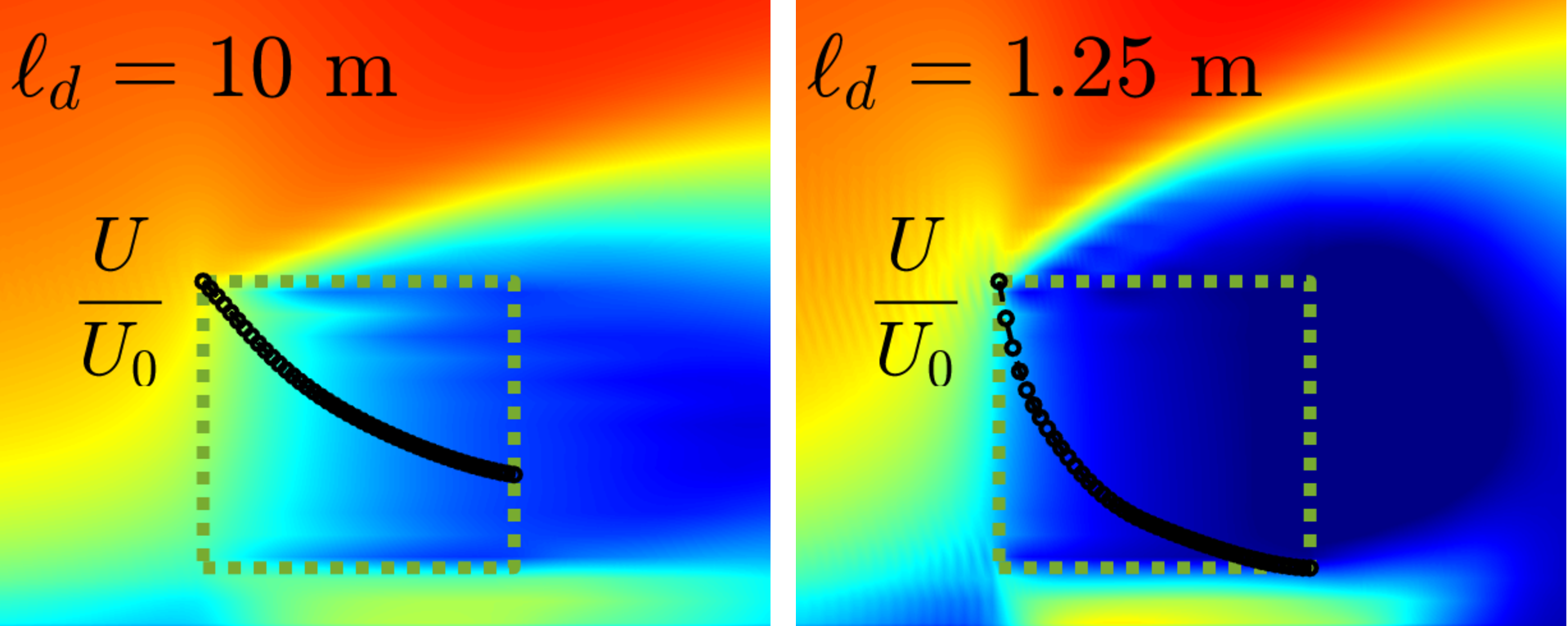}
\end{graphicalabstract}

\begin{highlights}

\item The drag properties of trees are determined by a single drag length $\ell_d = (a C_d^V)^{-1}$ where $a$ is the leaf-area density and $C_d^V$ is the volumetric drag coefficient.
\item Field and wind-tunnel measurements with real vegetation find the median drag length to be $21$ m for trees and $0.7$ m for low vegetation. Whereas median of $\ell_d$ values used by numerical models and wind-tunnel tree models (assuming geometric scaling $1:100$) is about $5$ m, suggesting possible overestimation of tree drag by these models.
\item In order to maintain the same dynamic conditions as at full scale, geometric scaling must be applied to the drag length of model trees in wind tunnels.
\item The aerodynamic porosity, often measured in wind-tunnel experiments provides direct access to the drag length $\ell_d$ and thus to $a C_D^V$. It is not crucial to have measurement of the actual drag force to determine $\ell_d$.

\end{highlights}


\end{frontmatter}

\begin{table*}[!t]   
    \begin{framed}
    
        \nomenclature{$a$}{Leaf-area density}
        
        \nomenclature{$A_{\rm F}$}{Projected windward frontal area of a tree canopy}
        
        \nomenclature{$c$}{Velocity shape coefficient, $\displaystyle c = {\langle \overline{u}^2 \rangle_{yz}}/{\langle \overline{u} \rangle_{yz}^2} = {\langle \overline{u}^2 \rangle_{yz}}/{U^2}$}
        
        \nomenclature{$C_d$}{Bulk drag coefficient based on projected frontal area, $\displaystyle C_d = {2 F_d}/{(\rho A_{\rm F} U_\infty^2)}$}
        
        \nomenclature{$C_d^V$}{Volumetric drag coefficient}
        
        \nomenclature{$Dr$}{Vegetation drag number, $\displaystyle Dr = H_b/\ell_d$}
        
        \nomenclature{$F_d$}{Bulk drag force}
        
        \nomenclature{$h$}{Height of a tree canopy along $z$-direction}
        
        \nomenclature{$h_0$}{Tree canopy base height from ground}
        
        \nomenclature{$H$}{Tree canopy crown height from ground, $H = h_0 + h$}
        
        \nomenclature{$H_b$}{Average building height}
        
        \nomenclature{$L$}{Length of a tree canopy along $x$-direction}
        
        \nomenclature{$p_{lw}$, $p_{ww}$}{Kinematic pressure at leeward and windward sides of the canopy, respectively}
        
        \nomenclature{$\overline{p}$}{Kinematic pressure, averaged over time}
        
        \nomenclature{$\langle \overline{p} \rangle_{yz}$}{Average horizontal pressure over the tree frontal area, $\displaystyle \langle \overline{p} \rangle_{yz} = A_F^{-1} {\iint_{A_{\rm F}} \overline p (x, y, z) \, \d y \, \d z}$}
        
        \nomenclature{$Re$}{Reynolds number based on the building height, $\displaystyle Re = \frac{U_{\infty} H_b}{\nu}$}
        
        \nomenclature{$s$}{Geometric scaling factor}
        
        \nomenclature{$\vec{S_u}$}{Volumetric source/sink term to model trees in CFD simulations, $\displaystyle \vec{S_u} = -aC_d^V \lvert \vec u \rvert \vec u$}
        
        \nomenclature{$S_u$}{Streamwise component of $\displaystyle \vec{S_u}$}
        
        \nomenclature{$\vec u$}{Wind velocity vector}
        
        \nomenclature{$\overline{u}$}{Streamwise component of wind velocity averaged over time}
        
        \nomenclature{$U$, $\langle \overline{u} \rangle_{yz}$}{Streamwise wind velocity averaged over the tree frontal area, $\displaystyle U  = \langle \overline{u} \rangle_{yz} = A_F^{-1} {\iint_{A_{\rm F}} \overline u (x, y, z) \, \d y \, \d z}$}
        
        \nomenclature{$U_0$, $U_L$}{$U$ at windward and leeward planes of a tree canopy, respectively, $\displaystyle U_0 = U\rvert_{x=0}$ and $\displaystyle U_L = U\rvert_{x=L}$}
        
        \nomenclature{$U_{\infty}$}{Free stream wind speed, \textit{i.e.}, $U$ at a far upstream location}
        
        \nomenclature{$V$}{Volume of a tree canopy}
        
        \nomenclature{$W$}{Width of a tree canopy along $y$-direction}
        
        \nomenclature{$x$, $y$, $z$}{Coordinate axes}
        
        \nomenclature{$\alpha$}{Aerodynamic porosity, $\displaystyle \alpha = U/U_0$}
        
        \nomenclature{$\alpha_L$}{Aerodynamic porosity of the entire tree canopy based on windward and leeward planes, $\displaystyle \alpha_L = \alpha\rvert_{x=L} = U_L/U_0$}
        
        \nomenclature{$\beta$}{A lengthscale coefficient assumed to be constant along the length of a tree canopy}

        \nomenclature{$\hat{\beta}$}{Prediction for $\beta$ based on regression model}
        
        \nomenclature{$\Delta p$}{Kinematic pressure difference between the windward and leeward sides of the canopy, $\Delta p = p_{ww} - p_{lw}$}
        
        \nomenclature{$\kappa$}{Factor relating bulk drag coefficient and aerodynamic porosity, $\displaystyle \kappa = \frac{C_d}{1-\alpha_L^2} = c(1+\beta)\left(U_0/U_{\infty}\right)^2$}
        
        \nomenclature{$\ell_d$}{Tree drag length, $\displaystyle \ell_d = \left( a C_d^V \right)^{-1}$}

        \nomenclature{$\hat{\ell}_d$}{Prediction for $\ell_d$ by making use of the regression model of $\beta$}
        
        \nomenclature{$\lambda$}{Pressure loss coefficient, $\lambda = 2\Delta p/(U^2 L)$}
        
        \nomenclature{$\nu$}{Kinematic viscosity of wind}
        
        \nomenclature{$\rho$}{Wind density}
        
        \printnomenclature
    \end{framed}
\end{table*}

\section{Introduction}
\label{sec:intro}

Trees are one of the central features of an urban environments. Trees make up at least 10\% of the total surface area in most cities and are often referred to as the urban forest~\citep{oke1989micrometeorology}. Trees make several contributions to the urban ecosystem, in particular flood water mitigation, reduction of heat stress through shading and evapotranspiration, improvement of air quality, and noise reduction~\citep{oke2017urban,bozovic2017blue}. Trees also play an important role in attenuating pedestrian-level winds~\citep{chen2021integrated}. Indeed, trees are regularly used as wind-breaks~\citep{smith2021windbreaks,weninger2021ecosystem}.

Wind engineering practitioners have increasingly adopted a computational fluid dynamics (CFD) model approach to model the pedestrian level wind environment and provide feedback on the thermal comfort and safety conditions for the public. Given their prevalence, it is critical to include vegetation in these simulations \citep{salim2015including}, which presents two distinct challenges. The first is to select an appropriate tree model given their complex interactions with the wind field, temperature, humidity, long- and shortwave radiation, and air quality~\citep[\textit{e.g.}][]{manickathan2018comparative, grylls2021tree}. The required complexity of the tree model will depend on the modelling needs of a project -- a wind study will require a less complex tree model than an urban microclimate study. The second challenge is -- once a suitable tree model has been selected -- what the appropriate parameter values are, as these will differ substantially depending on the characteristics of the tree (\textit{e.g.} species, age, season, etc.).

The drag exerted by trees is typically modelled, within CFD, as a volumetric sink term ($\vec {S_u}$) of momentum with a quadratic dependence on velocity~\citep{shaw1992large,raupach1996coherent,finnigan2000turbulence,nebenfuhr2015large},
\begin{equation}
\vec {S_u} = -a C_d^V \lvert \vec u \rvert \vec u \, ,
\label{eq:tree_source}
\end{equation}
where $\vec u$ denotes the wind velocity vector (typically the Reynolds-average), $a$ is the leaf-area density \citep{oke2017urban}, and $C_d^V$ is the volumetric drag coefficient. Models that are based on the Reynolds-averaged Navier-Stokes equations typically also add terms in the turbulence kinetic energy and dissipation equations~\citep{sanz2003note,mochida2008examining,salim2015including,buccolieri2018review}, but importantly these also include $a$ and $C_d^V$.

\begin{figure}
\centering
\includegraphics[scale=0.262]{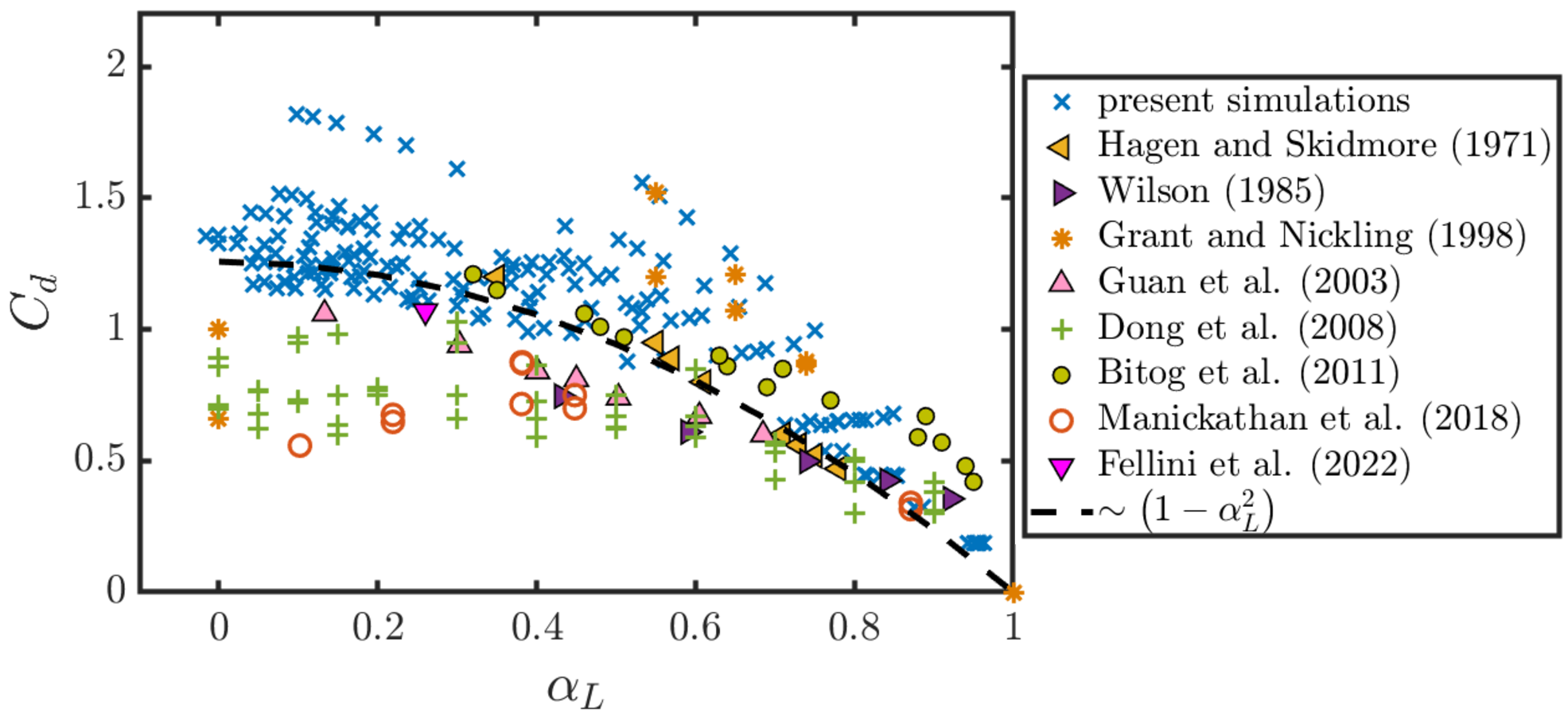}
\caption{Literature data reflecting overall $C_d \sim \left(1-\alpha_L^2\right)$ behaviour corroborates the proposed model.}
\label{fig:Cd_vs_aeroporo}
\end{figure}

Two distinct tree drag regimes can be identified: the extended canopy regime and the wind-break regime. The first is the drag exerted onto the atmosphere by extended forests, where the tree drag is in equilibrium with the vertical divergence of the vertical transport of horizontal momentum by turbulence~\citep{belcher2008flows}. As $a$ is formally defined as the one-sided leaf area per unit volume of air, this allows for a straightforward calculation of $C_d^V$. The typical values for both $a$ and $C_d^V$ from literature are listed in Table~\ref{tab:literature values 1}.
The second drag regime comprises wind-breaks which are much shorter~\citep{lyu2020review}, \textit{e.g.} a tree line or a small park, where the momentum from incoming wind is reduced by the tree drag. In this case, one often relies on wind-tunnel studies that are able to determine the drag force at high accuracy; after which it is straightforward to calculate a drag coefficient as \citep{mayhead1973some,guan2003wind,cullen2005trees}
\begin{equation}
C_d = \frac{2 F_d}{\rho A_{\rm F} U_\infty^2} \, ,
\label{eq:CDdef}
\end{equation}
where $F_d$ is the drag force, $A_{\rm F}$ is the projected frontal area of the tree canopy facing the incoming wind, $U_{\infty}$ indicates the average wind speed at far upstream, and $\rho$ indicates air density. 
There are two issues with representing trees using $C_d$. The first is that the value of $C_d$ is not constant (Table~\ref{tab:literature values 2}) but has a strong dependence on the aerodynamic porosity \citep{salim2015including}
\begin{equation}
 \alpha(x) = \frac{\iint_{A_{\rm F}} \overline u (x, y, z) \d y \d z}{\iint_{A_{\rm F}} \overline u (0, y, z) \d y \d z} \, .
 \label{eq:alphadef}
\end{equation}
Here it is assumed that the wind velocity is aligned with the $x$-direction, and the wind-break starts at $x=0$. Note that $\alpha$ is a function of $x$ which implies that there is no universal value of $C_d$ for trees as is often observed for high Reynolds number bluff body flows. Often wind-tunnel studies report $\alpha_L \, (= \alpha\rvert_{x=L})$ which is the ratio of the volume flux on the windward and leeward side of the wind-break~\citep{guan2003wind}. Figure~\ref{fig:Cd_vs_aeroporo} shows the variation of $C_d$ with respect to $\alpha_L$ reported in literature, along with values obtained from the simulations carried out in this study. There are empirical models that link $C_d$ and $\alpha_L$~\citep{grant1998direct,de2000wind,guan2003wind} but these models lack a theoretical basis.
The second issue of representing tree drag with Eq.~\eqref{eq:CDdef} is that it is not clear how the bluff body drag coefficient $C_d$ is related to the volumetric drag coefficient $C_d^V$. Alternatively, \citet{gromke2016influence,gromke2018wind} used a pressure loss coefficient (defined as the static pressure loss per canopy length normalized by the dynamic pressure) to describe the drag induced by the porous foam tree models used in their wind-tunnel experiments. Again, it is not clear how this pressure loss coefficient relates to $a$ and $C_d^V$.

The accepted value of $C_d^V$ and $a$ in a forest canopy can vary depending on several factors such as tree species, size, age, canopy density, seasonal foliage shedding, and wind conditions. Table~\ref{tab:literature values 1} summarises the widely reported values of $a$ and $C_d^V$; field and wind-tunnel experiments of real trees indicate that the drag coefficient largely varies within $0.1 \le C_d^V \le 3.0$ and the leaf-area density approximately ranges between $0.1 < a < 1.6$ m$^{-1}$ for trees and $1.0 < a < 12.6$ m$^{-1}$ for low vegetation (\textit{e.g.} crops). Although $a$ and $C_d^V$ both varies along the height of the tree foliage, constant values are commonly taken in CFD simulations~\citep{buccolieri2018review,fu2024should}. Instead of species specific values, majority of literature considers $C_d^V = 0.2$ to reflect an average value for urban vegetation in CFD modelling, and varied the leaf-area density to incorporate seasonal effects; see Table~\ref{tab:literature values 1}.

As per the tree model in Eq.~\eqref{eq:tree_source}, CFD simulations with a tree canopy require $a$ and $C_d^V$ as the main input parameters to model the tree. However, wind-tunnel experiments with model trees typically measure the quantities $C_d$ and $\alpha_L$. A direct correlation between $a$, $C_d^V$ and $C_d$, $\alpha_L$ is yet to be achieved. Since $C_d$ is defined based on a two-dimension projected area, is fundamentally different from the volumetric coefficient $C_d^V$. Thus it is not clear what values for $a$ and $C_d^V$ to use as inputs to replicate the same aerodynamic characteristics of an experimental model tree in a corresponding CFD simulation accurately. The present study aims to establish a connection between the two set of parameters in an appropriate manner such that both the CFD and wind-tunnel model trees yield comparable aerodynamic traits. The work will provide an insight into the drag caused by trees on the air flow, and will propose the appropriate parameter values for use in CFD, and how to infer these from wind-tunnel and field data. 
In order to do so, we construct a simple analytical model for the wind flow inside trees in the wind-break regime (Section~\ref{sec:analytical_model}) and conduct a parametric study using large-eddy simulations (LES) for wind-breaks having various geometric configurations (Section~\ref{sec:simulations}). Results are shown in Section~\ref{sec:results} and we employ a nonlinear regression to fit a lengthscale parameter in the analytical model. The implications of our findings are discussed in Section~\ref{sec:implecations} and conclusions are made in Section~\ref{sec:con}.

\section{Analytical model for flow inside the tree canopy}
\label{sec:analytical_model}

\begin{figure}
\centering
\includegraphics[scale=0.35]{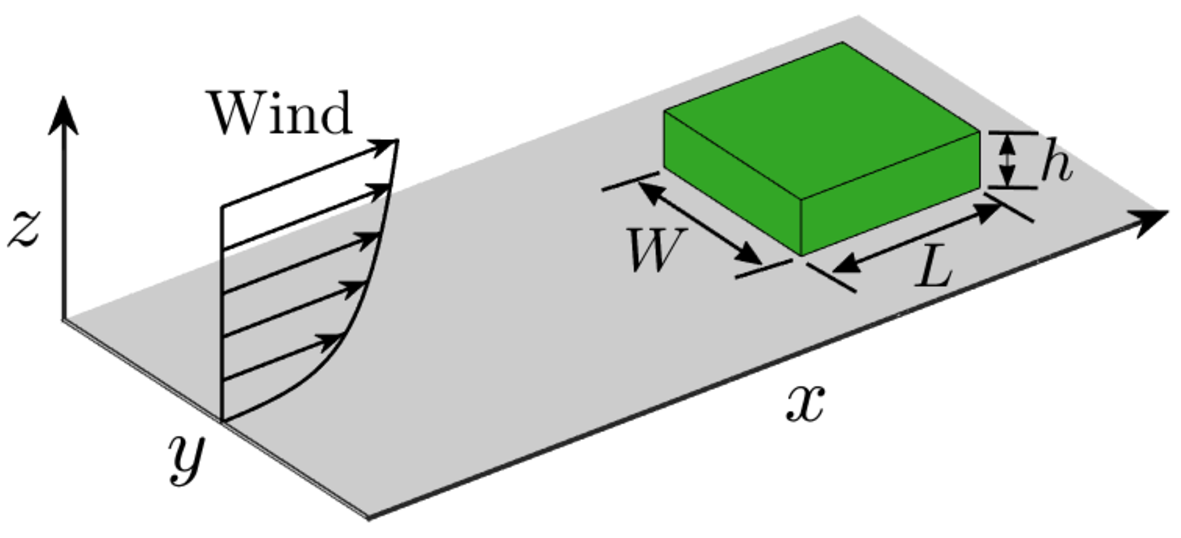}
\caption{Schematic of a tree canopy.}
\label{fig:schematic_canopy}
\end{figure}

Consider a homogeneous vegetation canopy of length $L$, width $W$ and height $h$, with the canopy base at height $h_0$ (Fig.~\ref{fig:schematic_canopy}). 
The canopy starts at $x=0$ where $x$ is in the direction of the wind. The inflow profile is turbulent. The aim is to  create an analytical model for the flow inside the canopy. 

The starting point of the derivation is a simplified Reynolds-averaged horizontal momentum equation that only considers horizontal advection, the drag exerted by the tree on the fluid, and the pressure gradient:
\begin{equation}
    \overline{u}\frac{\partial \overline{u}}{\partial x} + \frac{\partial \overline{p}}{\partial x} = 
    -aC_d^V \overline{u}^2 \, .
    \label{eq:momu_tree_6}
\end{equation}
Here, $u$ is the wind velocity in $x$-direction, and $p$ is the kinematic pressure; the overbar sign ( $\displaystyle \overline{ * }$ ) indicates the temporal mean.
Averaging Eq.~\eqref{eq:momu_tree_6} over the canopy height and width, and introducing a tree drag length $\ell_d = (a C_d^V)^{-1}$, one obtains
\begin{equation}
    \frac{\d \langle \overline{u}^2 \rangle_{yz}}{\d x} 
    + 2\frac{\d \langle \overline{p} \rangle_{yz}}{dx} = 
    -\frac{2}{\ell_d} \langle \overline{u}^2 \rangle_{yz} 
    \, ,
    \label{eq:momu_tree_av_6}
\end{equation}
where $\displaystyle \langle * \rangle_{yz} = (Wh)^{-1}\int_{h_0}^{h_0+h} \int_{-W/2}^{W/2} (\, * \,) \; \d y \, \d z$.
Note that $\ell_d$ fully characterises the drag properties of the vegetation; this lengthscale was found to be the central parameter for the adjustment length in inhomogeneous vegetation canopies \citep{finnigan1995turbulent, finnigan2000turbulence,banerjee2013mean}.

\begin{figure}
\centering
\includegraphics[scale=0.55]{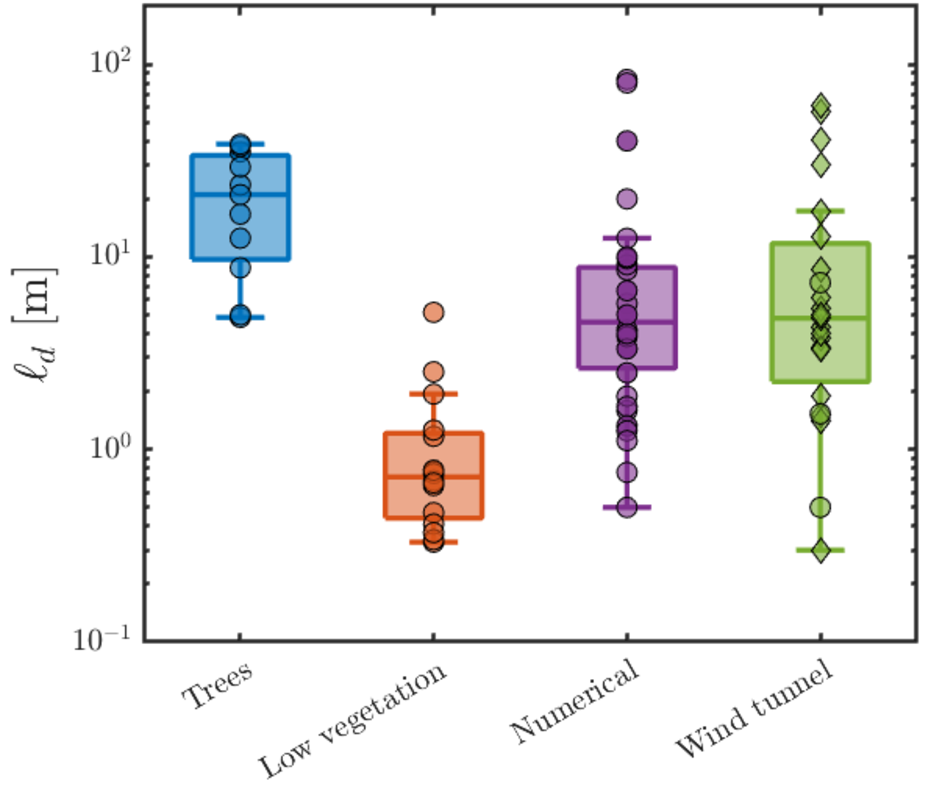}
\caption{Box-plot of drag length variation for trees, low vegetation, numerical-model trees, and wind-tunnel model trees. Markers indicate the individual data point from Tables~\ref{tab:literature values 1} and~\ref{tab:literature values 2}. The circles indicates full-scale values based on literature and the diamonds represent drag lengths of the wind-tunnel model trees in Table~\ref{tab:literature values 2} that have been converted to full-scale values assuming a scaling factor $s=100$.}
\label{fig:elld_box}
\end{figure}

Typical $\ell_d$ values of vegetation canopies as reported by various field- studies, wind-tunnel studies, and simulations in literature are summarised in Table~\ref{tab:literature values 1} and are displayed in Fig.~\ref{fig:elld_box} as box-plots. Here, we have distinguished between tree canopies and low-vegetation canopies (\textit{i.e.} crops). The data shows that the median value of $\ell_d$ for trees is $21$ m, with an interquartile range of $10 < \ell_d < 34$ m. Low vegetation has a median value of $\ell_d=0.7$ m with an interquartile range $0.4 < \ell_d < 1.2$ m. Also shown in the figure are the values used in numerical models which show a median value of $4.6$ m with with an interquartile range of $2.6 < \ell_d < 8.8$ m. We note that this is substantially smaller than the field studies suggest, which in turn might overestimate the effect of trees on wind mitigation. However, note that it is unclear how much the trunks contribute to the drag in the field studies, since trunks are likely to be less important for extended canopies (which are what most of the field studies report on) than for wind-breaks.
The figure also shows estimates for wind-tunnel model trees (green colour); the calculation method and results will be discussed in Sections~\ref{sec:interlink} and~\ref{sec:wt scaling}.

To obtain a solution for Eq.~\eqref{eq:momu_tree_av_6}, we define the average velocity $U=\langle \overline{u} \rangle_{yz}$, average pressure $P=\langle \overline{p} \rangle_{yz}$ and a shape coefficient $c=\langle \overline{u}^2 \rangle_{yz} / U^2$ which is assumed to be independent of $x$. With these definitions, Eq.~\eqref{eq:momu_tree_av_6} is given by
\begin{equation}
    \frac{\d U^2}{\d x} 
    + \frac{2}{c}\frac{\d P}{dx}
    = -\frac{2}{\ell_d} U^2  \, .
    \label{eq:momu_tree_ode_6}
\end{equation}
We assume that the pressure gradient has the same form as the inertial term inside the tree canopy, \textit{i.e.},
\begin{equation}
\frac{2}{c}\frac{\d P}{dx} = \beta \frac{\d U^2}{\d x} \, .
\end{equation}
This implies that the pressure inside the canopy is assumed to evolve as $\displaystyle P = P_0 + \frac{1}{2}\beta c (U^2-U_0^2)$, where $P_0 = P \rvert_{x=0}$, $U_0 = U\rvert_{x=0}$, and $\beta$ is a coefficient assumed to be independent of $x$. The results will show that this assumption works reasonably well. Although other parameterisations are conceivable, this one stands out for its simplicity. It should be recognised that the coefficient $\beta$ represents, apart from the effect of pressure, all three-dimensional effects not included in Eq.~\eqref{eq:momu_tree_6}. With this assumption, Eq.~\eqref{eq:momu_tree_ode_6} becomes
\begin{equation}
    (1+\beta) \frac{\d U^2}{\d x}
    = -\frac{2}{\ell_d} U^2. 
    \label{eq:momu_tree_ode_a_6}
\end{equation}
Together with the boundary condition $U(x=0) = U_0$, Eq.~\eqref{eq:momu_tree_ode_a_6} has the solution
\begin{equation}
U =  U_0 \, \exp\left(- \frac{x}{(1+\beta) \ell_d} \right) \, ,
   \label{eq:U2_6}
\end{equation} 
which implies that the aerodynamic porosity $\alpha$ given in Eq.~\eqref{eq:alphadef} evolves as 
\begin{equation}
\displaystyle    \alpha = \frac{U}{U_0} = \exp \left( - \frac{x}{(1+\beta) \ell_d} \right) \, .
   \label{eq:alpha2_6}
\end{equation}
Small values of $x/\ell_d \ll 1$ are representative of the wind-break regime, and large values of $x/\ell_d \gg 1$ of the continuous canopy regime (where the model Eq.~\eqref{eq:momu_tree_6} is not valid). Assuming that the wind-break regime is present until $U$ drops to $5$\% of $U_0$, the associated canopy length is $3 (1+\beta)\ell_d$.

The canopy drag force is given by
\begin{equation}
\begin{split}
    F_D &= \rho  \int_V S_u \d V
= \frac{\rho W h c}{\ell_d} \int_0^L U^2 \, \d x 
\\ &= \frac{1}{2} \, \rho W h c U_0^2 (1+\beta) \left\{1 - \exp\left(- \frac{2L}{(1+\beta)\ell_d}\right) \right\}
.
\end{split}
\label{eq:drag 6}
\end{equation}
Therefore, the drag coefficient defined in Eq.~\eqref{eq:CDdef} is given by 
\begin{equation}
    C_d
    = c(1+\beta)\left(\frac{U_0}{U_{\infty}}\right)^2 \, \left\{1 - \exp\left(- \frac{2L}{(1+\beta)\ell_d}\right) \right\}.
\label{eq:Cd-CdV_6}
\end{equation}
Note that $C_d$ attains $95\%$ of its wind-break value in $1.5(1+\beta)\ell_d$ rather than $3 (1+\beta)\ell_d$, due to its dependence on $U^2$. Finally using Eq.~\eqref{eq:alpha2_6}, $C_d$ can be directly related to the aerodynamic porosity as
\begin{equation}
    C_d = \kappa \, \left(1 -\alpha_L^2 \right),
    \label{eq:Cd-alpha}
\end{equation} 
where $\kappa=c(1+\beta)\left(U_0/U_{\infty}\right)^2$, and $\alpha_L = \alpha\rvert_{x=L}$. This relation explains why the literature~\citep{hagen1971windbreak,wilson1985numerical,grant1998direct,dong2008analysis} finds such a strong relationship between $C_d$ and $\alpha$; see Fig.~\ref{fig:Cd_vs_aeroporo} which summarises the spread in $C_d$ of wind-breaks and trees in relation to $\alpha$ as reported in literature and obtained in the simulations carried out in this work. A notable spread in $C_d$ for a given porosity is due to different tree canopy types, and also because the method of calculating the aerodynamic porosity varies across the listed studies. According to Eq.~\eqref{eq:Cd-alpha}, $C_d$ starts at $0$ when $\alpha_L = 1$, and then increases as $\sim (1-\alpha_L^2)$ with decreasing $\alpha_L$. Eventually, $C_d$ reaches a value of $\kappa$ at $\alpha_L = 0$. These outcomes corroborate well the physical flow behaviour inside a tree canopy as delineated in the literature. \citet{dong2007wind} reported a critical value of $\alpha_L$ to be $0.3$. Above this critical $\alpha_L$, there exists a dominant bleed flow through the tree causing the drag force to decrease quickly with increasing $\alpha_L$. Below critical $\alpha_L$, there exists only little bleed flow through the tree and most portion of the incoming wind flows around, resulting in a recirculation zone downstream~\citep{manickathan2018comparative}.

In the simplified canopy model proposed here, it is assumed that the cross-section area of the tree canopy remains constant both along its height and length. Moreover, the leaf-area density $a$ is assumed to be constant along the height, and the effect of foliage reconfiguration~\citep{rudnicki2004wind,vollsinger2005wind,manickathan2018comparative} has been disregarded. These may not be always true for real trees in nature. However, the importance of the proposed theoretical model is that it provides an explicit understanding of the bulk (average) flow behaviour inside the tree canopy. Certainly, the theory can be extended in future taking the spatial variation of the tree properties into consideration. Despite these obvious limitations, the present model provides interesting insights into the determination of the volumetric drag coefficient $C_d^V$: 
\begin{enumerate}
\item[i.] Aerodynamic porosity is a proxy for $\ell_d$, since $ \displaystyle \alpha_L = \alpha\rvert_{x=L} = \exp\left(-L/(1+\beta)\ell_d\right)$. Thus the aerodynamic porosity provides direct access to the drag length $\ell_d$. Note that it is unnecessary to determine the actual drag force, although that will of course provide another independent estimate of $\ell_d$. 
\item[ii.] As far as total drag is concerned, the leaf-area density and the volumetric drag coefficient are exchangeable and combine into a single drag length $\ell_d = (a C_d^V)^{-1}$. The corollary is that $a$ must be determined independently; once this is done, the appropriate value for $C_d^V$ can be obtained using $\ell_d$.
\item[iii.] A limitation of the current model is that $\alpha$ is a meaningful quantity only in the wind-break regime as it decays to zero for extended canopies. Measurements of longer canopies do not necessarily provide better information if the canopy length is longer than the wind-break regime $3 (1+\beta) \ell_d$.  
\end{enumerate}

Equations~\eqref{eq:alpha2_6} and \eqref{eq:Cd-alpha} model the overall aerodynamic traits of the vegetation canopy. The former indicates how the bulk flow evolves inside the canopy whereas the latter establishes a direct correlation between $\alpha$ and $C_d$. However, the coefficient $\beta$ in Eq.~\eqref{eq:alpha2_6} is unclosed and there is insufficient information available in literature to determine it. A combination of both recirculation and bleed flow affects the value of $\beta$, and it is likely to depend on $L, \, W, \, H, \, \ell_d, \, U_{\infty}, \, U_0$ etc. Additionally, the dominant three-dimensional nature of the flow at low $\ell_d$ (high $aC_d^V$) can significantly influence $\beta$, as the momentum balance in such situations occurs in the vertical direction instead of the horizontal momentum balance considered in Eq.~\eqref{eq:momu_tree_6}. In order to provide a suitable parameterisation of $\beta$, LES is used to carry out a wide parametric study.

\section{Simulation details}
\label{sec:simulations}

The simulations are performed using uDALES~\citep{suter2022udales,owens2024conservative} which is a multi-physics micro-climate modelling framework for the urban environment. It performs LES of the incompressible Navier-Stokes equations within the Boussinesq approximation. The effects of urban surfaces are taken into account in terms of a novel conservative immersed boundary method~\citep{owens2024conservative}. uDALES uses wall functions to model surface fluxes (\textit{e.g.} shear stress) which are then converted to appropriate source/sink terms to apply in the momentum equations. \citet{grylls2021tree} included trees in the uDALES framework, modelling the drag, shading, evaporation and deposition. The primary focus in the current work is the drag behaviour under neutral atmospheric conditions. Trees are modeled in uDALES as rectangular blocks, and the source/sink term defined in Eq.~\eqref{eq:tree_source} is applied to the grid points that fall within the volumetric tree blocks. The turbulence is resolved up to the grid scale and the subgrid-scale turbulence is modelled using the Vreman eddy viscosity~\citep{vreman2004eddy}. More details on the uDALES framework and the tree modelling can be found in the works of \cite{grylls2021tree,owens2024conservative}.

\begin{figure}
\centering
\includegraphics[scale=0.3]{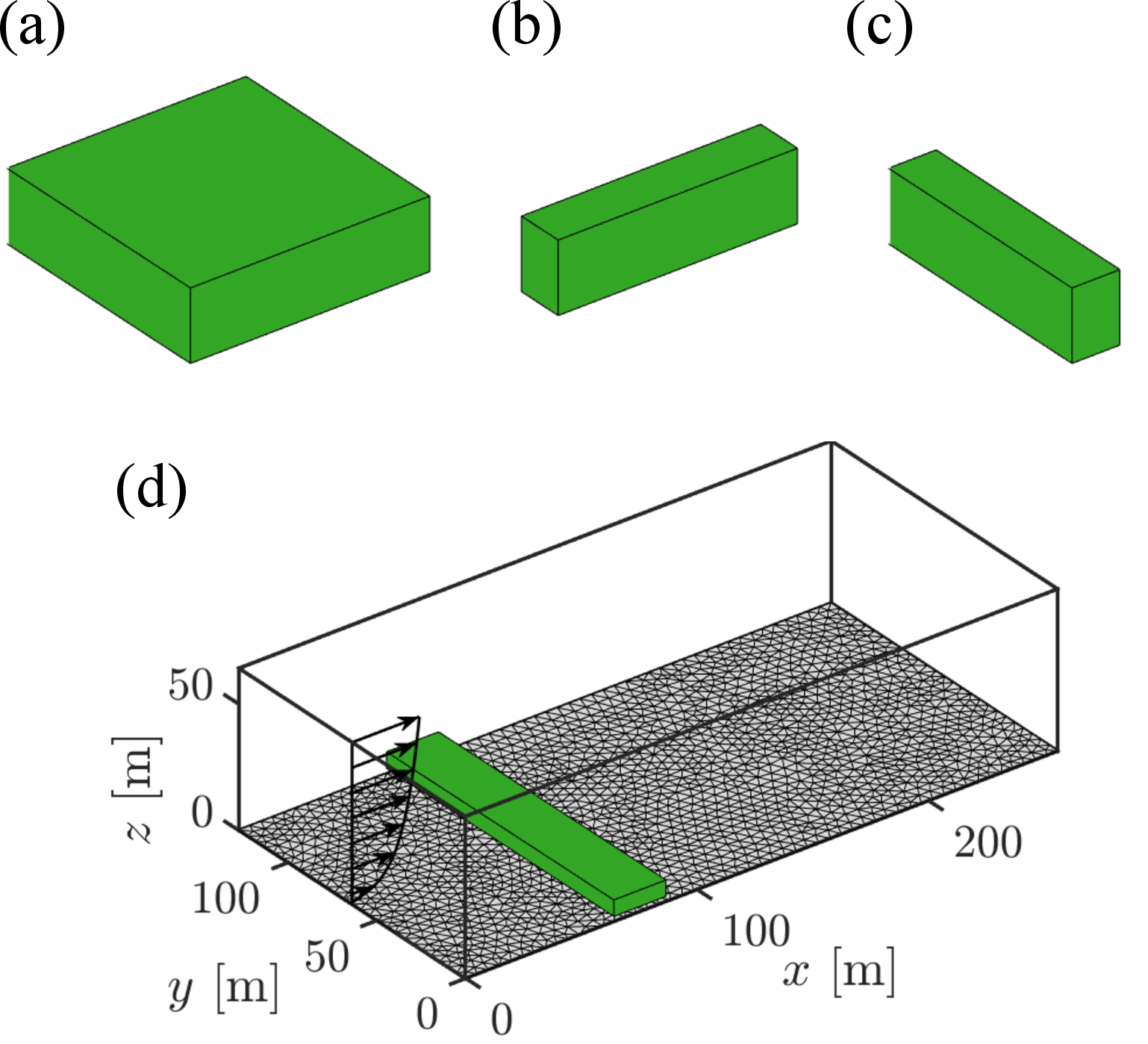}
\caption{Schematic tree canopy blocks: a) square (S), b) long (L), c) wide (W), and d) infinitely wide (I) tree canopy along with the computational domain.}
\label{fig:tree_domain}
\end{figure}
 
The default tree geometry is chosen to be a cuboid of size $L \times W \times h = 4.5 \times 4.5 \times 6.5$ m$^3$, taking inspiration from the wind-tunnel experiments of \citet{fellini2022high}, resembling the canopy of a single tree. The base of the tree is at $h_0=2.0$ m height from the ground, hence the crown of the tree is at a height of $H = h_0+h = 8.5$ m from the ground. Starting from this default tree geometry, a large parametric study is set up, with canopies categorised as:
\begin{enumerate}
    \item[i.] square (S): five such cases with $L=W=4.5$ m, $9.0$ m, $13.5$ m, $18.0$ m and $22.5$ m; 
    \item[ii.] long (L): four such cases with $L=9.0$ m, $13.5$ m, $18.0$ m and $22.5$ m, keeping $W = 4.5$ m fixed; 
    \item[iii.] wide (W): four such cases with $W=9.0$ m, $13.5$ m, $18.0$ m and $22.5$ m, keeping $L = 4.5$ m fixed; 
    \item[iv.] infinitely wide tree canopy (I): here the canopy width spanned the entire computational domain width emulating an infinitely wide tree canopy, two such cases with $L = 4.5$ m and $22.5$ m are considered. 
\end{enumerate}

\begin{table*}[t!]
		\caption{Summary of the total $168$ different simulation cases.}
		\label{tab:simulation cases}
 \small
	\begin{center}
		\begin{tabular}{p{1.5cm}|p{2.5cm}|p{8cm}}
			\textbf{Type} & \textbf{Canopy size} & \textbf{Canopy length / Drag length} \\
			 &  $L\times W\times h$ [m$^3$] & $L/\ell_d$ [-] \\
\hline
\hline
             S45 & $4.5 \times 4.5 \times 6.5$ & $0.09, \quad 0.45, \quad 0.90, \quad 1.35, \quad 1.80, \quad 2.70, \quad 3.60, \quad 4.50$ \\
             S90 & $9.0 \times 9.0 \times 6.5$ & $0.18, \quad 0.90, \quad 1.80, \quad 2.70, \quad 3.60, \quad 5.40, \quad 7.20, \quad 9.00$ \\
             S135 & $13.5 \times 13.5 \times 6.5$ & $0.27, \quad 1.35, \quad 2.70, \quad 4.05, \quad 5.40, \quad 8.10, \quad 10.8, \quad 13.5$ \\
             S180 & $18.0 \times 18.0 \times 6.5$ & $0.36, \quad 1.80, \quad 3.60, \quad 5.40, \quad 7.20, \quad 10.8, \quad 14.4, \quad 18.0$ \\
             S225 & $22.5 \times 22.5 \times 6.5$ & $0.45, \quad 2.25, \quad 4.50, \quad 6.75, \quad 9.00, \quad 13.5, \quad 18.0, \quad 22.5$ \\
\hline
             L90 & $9.0 \times 4.5 \times 6.5$ & $0.18, \quad 0.90, \quad 1.80, \quad 2.70, \quad 3.60, \quad 5.40, \quad 7.20, \quad 9.00$ \\
             L135 & $13.5 \times 4.5 \times 6.5$ & $0.27, \quad 1.35, \quad 2.70, \quad 4.05, \quad 5.40, \quad 8.10, \quad 10.8, \quad 13.5$ \\
             L180 & $18.0 \times 4.5 \times 6.5$ & $0.36, \quad 1.80, \quad 3.60, \quad 5.40, \quad 7.20, \quad 10.8, \quad 14.4, \quad 18.0$ \\
             L225 & $22.5 \times 4.5 \times 6.5$ & $0.45, \quad 2.25, \quad 4.50, \quad 6.75, \quad 9.00, \quad 13.5, \quad 18.0, \quad 22.5$ \\
\hline
             W90 & $4.5 \times 9.0 \times 6.5$ & $0.09, \quad 0.45, \quad 0.90, \quad 1.35, \quad 1.80, \quad 2.70, \quad 3.60, \quad 4.50$ \\
             W135 & $4.5 \times 13.5 \times 6.5$ & $0.09, \quad 0.45, \quad 0.90, \quad 1.35, \quad 1.80, \quad 2.70, \quad 3.60, \quad 4.50$ \\
             W180 & $4.5 \times 18.0 \times 6.5$ & $0.09, \quad 0.45, \quad 0.90, \quad 1.35, \quad 1.80, \quad 2.70, \quad 3.60, \quad 4.50$ \\
             W225 & $4.5 \times 22.5 \times 6.5$ & $0.09, \quad 0.45, \quad 0.90, \quad 1.35, \quad 1.80, \quad 2.70, \quad 3.60, \quad 4.50$ \\
\hline
             I45 & $4.5 \times 128 \times 6.5$ & $0.09, \quad 0.45, \quad 0.90, \quad 1.35, \quad 1.80, \quad 2.70, \quad 3.60, \quad 4.50$ \\
             I225 & $22.5 \times 128 \times 6.5$ & $0.45, \quad 2.25, \quad 4.50, \quad 6.75, \quad 9.00, \quad 13.5, \quad 18.0, \quad 22.5$ \\
\hline
             S225H135 & $22.5 \times 22.5 \times 13.5$ & $0.45, \quad 2.25, \quad 4.50, \quad 6.75, \quad 9.00, \quad 13.5, \quad 18.0, \quad 22.5$ \\
             S225H180 & $22.5 \times 22.5 \times 18.0$ & $0.45, \quad 2.25, \quad 4.50, \quad 6.75, \quad 9.00, \quad 13.5, \quad 18.0, \quad 22.5$ \\
             S135H135 & $13.5 \times 13.5 \times 13.5$ & $0.27, \quad 1.35, \quad 2.70, \quad 4.05, \quad 5.40, \quad 8.10, \quad 10.8, \quad 13.5$ \\
             S135H180 & $13.5 \times 13.5 \times 18.0$ & $0.27, \quad 1.35, \quad 2.70, \quad 4.05, \quad 5.40, \quad 8.10, \quad 10.8, \quad 13.5$ \\
             L135H135 & $13.5 \times 4.5 \times 13.5$ & $0.27, \quad 1.35, \quad 2.70, \quad 4.05, \quad 5.40, \quad 8.10, \quad 10.8, \quad 13.5$ \\
             W135H135 & $4.5 \times 13.5 \times 13.5$ & $0.09, \quad 0.45, \quad 0.90, \quad 1.35, \quad 1.80, \quad 2.70, \quad 3.60, \quad 4.50$ \\
		\end{tabular}
	\end{center}
\end{table*}

Unless specifically mentioned, the height of the tree canopy is kept constant as $h=6.5$ m. Additionally, a few selected cases are also run with different canopy heights, $h = 13.5$ m and $18.0$ m. For each of the above canopy dimensions, $aC_d^V$ is varied such that $aC_d^V \in \left[0.02, \; 0.1, \; 0.2, \; 0.3, \; 0.4, \; 0.6, \; 0.8, \; 1.0\right]$ m$^{-1}$, \textit{i.e.}, $\ell_d \in \left[50.0, \; 10.0, \; 5.0, \; 3.33, \; 2.5, \; 1.67, \; 1.25, \; 1.0\right]$ m, respectively. These values of $aC_d^V$ are selected in a way that they cover the spread of $C_d^V$ and $a$ reported in literature for various types of tree canopies as listed in Table~\ref{tab:literature values 1}. All simulation cases are systematically summarised in Table~\ref{tab:simulation cases} for clarity. 

\begin{figure}
\centering
\includegraphics[scale=0.32]{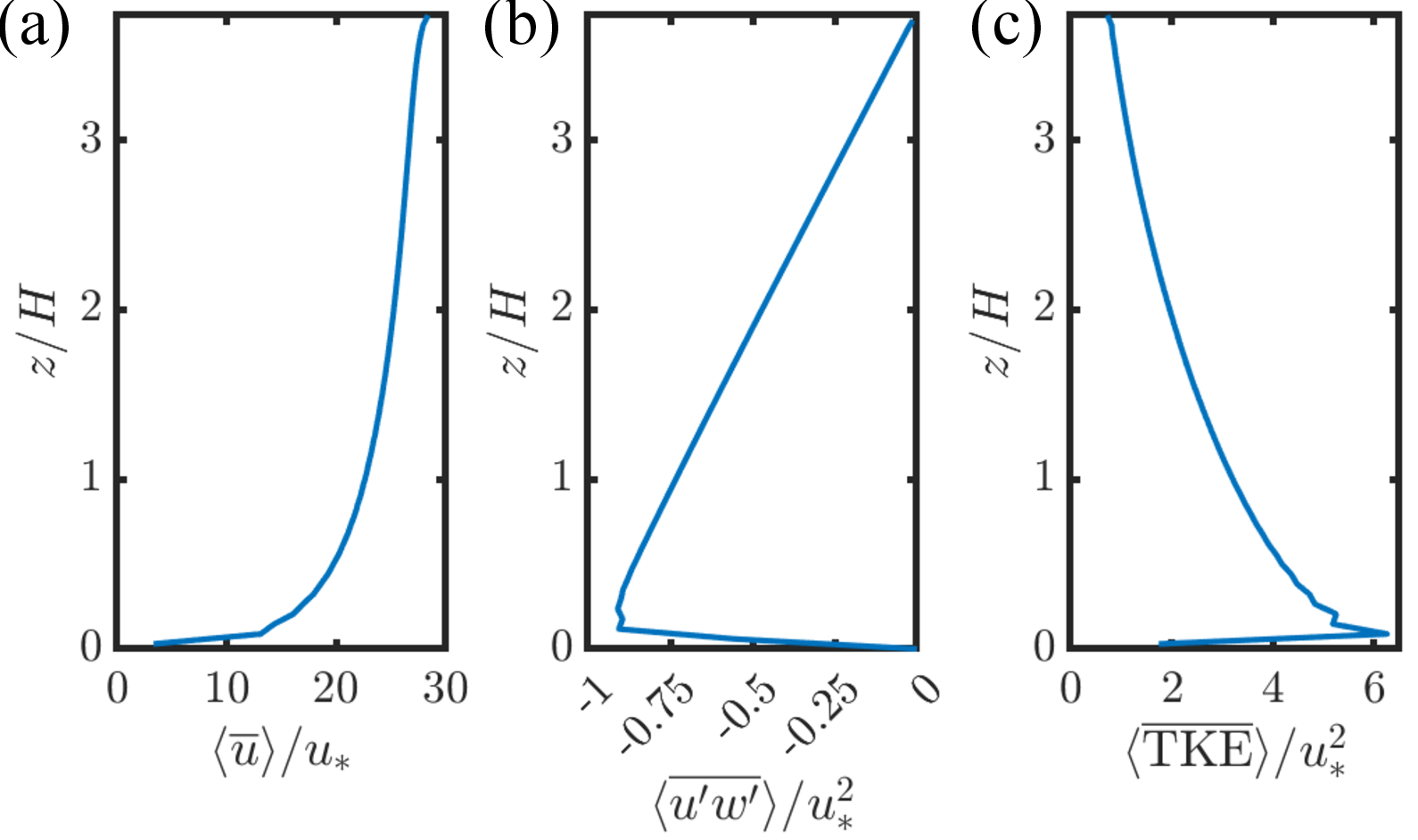}
\caption{Temporal and spatially averaged profile of the flow-field supplied at the inlet of the simulations with tree canopy; (a) stream-wise velocity, (b) Reynolds stress term, and (c) turbulent kinetic energy. $u_*$: shear/friction velocity.}
\label{fig:driver}
\end{figure}

The dimension of the computational domain is $256 \times 128 \times 32$ m$^3$, consisting of total $512 \times 256 \times 64$ grid points with approximately $0.5$ m uniform grid spacing in all directions. The tree blocks are placed at $64.0$ m distance from the inlet, and symmetrically along the $y$-direction. Schematic diagrams of the computational domain with different types of tree canopies are shown in Fig.~\ref{fig:tree_domain}. 
The computational domain size is such that the blockage ratio ($= ({\rm tree \; frontal \; area} / {\rm domain \; frontal \; area}) \times 100 \%$) remains below $5\%$ for the `S', `L' and `W' cases. 
The `I' cases and the taller canopy cases are run with a larger domain of $256 \times 128 \times 64$ m$^3$ ($512 \times 256 \times 128$ grid points) to maintain the acceptable limit of $10\%$ blockage ratio. Obviously, the tree canopies being porous in nature, flow can pass through them, as a result, the actual blockage ratio is less than what one obtains based on the frontal area.
Unless specifically mentioned, a prescribed inflow and a convective outflow boundary conditions are used in the $x$-direction, while the $y$-direction is periodic, and the top is free-slip. At the bottom, no-slip boundary condition is used with a flat ground surface having roughness length $0.05$ m. A precursor (driver) simulation~\citep{suter2022udales,owens2024conservative} with an empty domain is carried out first to generate a neutral turbulent atmospheric boundary layer which is then provided as inflow to the target simulations with tree canopy. 
The driver simulation is performed with an assigned constant volume flow rate forcing in the $x$-direction enforcing a bulk wind speed of $1.0$ m/s, and with periodic boundary condition in both $x$ and $y$-directions. 
Temporally and spatially averaged stream-wise velocity, Reynolds stress and turbulent kinetic energy profiles obtained from the driver simulation output are shown in Fig.~\ref{fig:driver}. This output velocity data obtained from the driver simulation is provided as inlet in the target simulations with the tree canopy.

\section{Results}\label{sec:results}

\begin{figure*}[t!]
\centering
\includegraphics[scale=0.35]{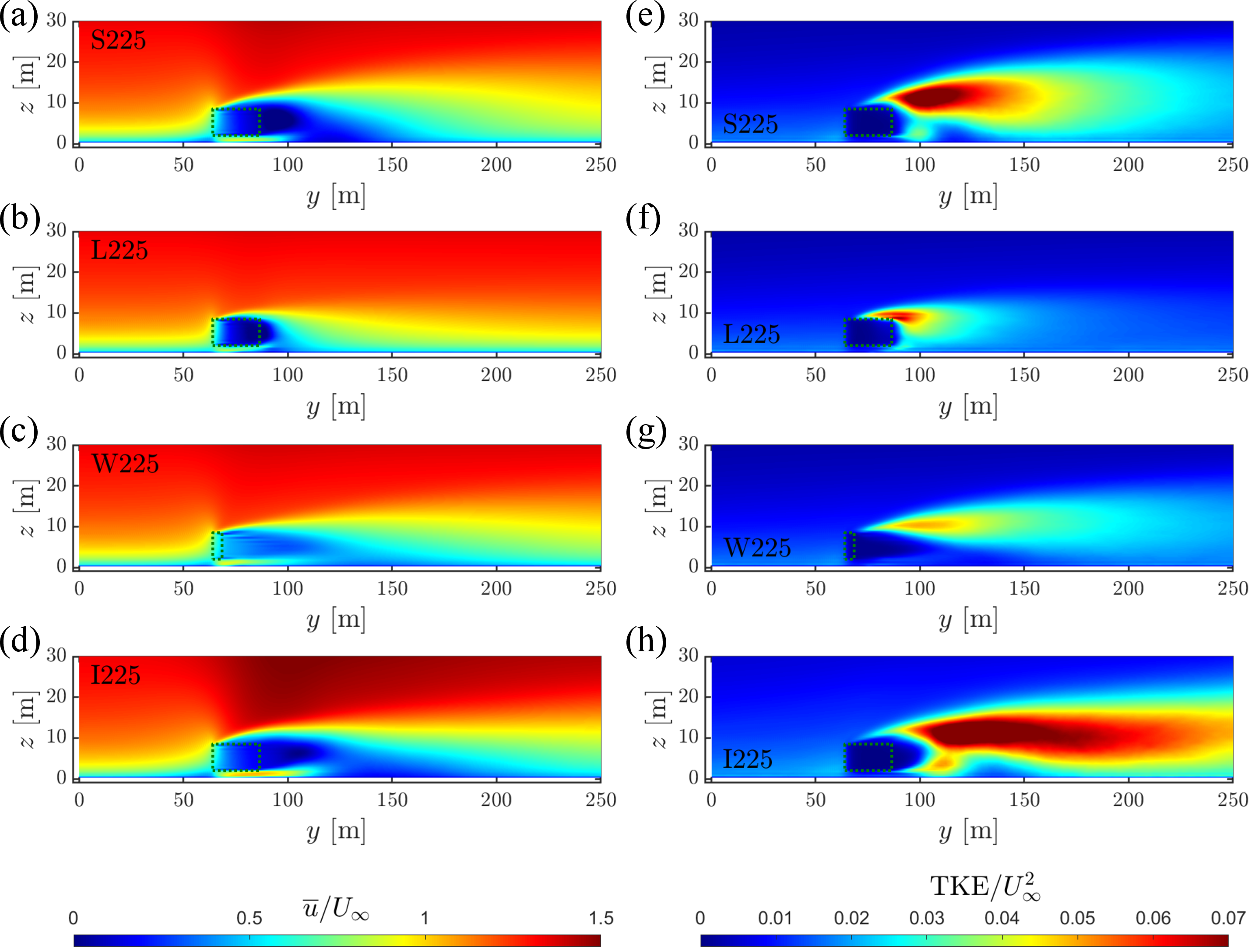}
\caption{Contour of mean (a-d) streamwise velocity and (e-h) turbulent kinetic energy at the mid-span vertical plane at $\ell_d = 3.33$ m for a representative case from each canopy type.}
\label{fig:u_and_tke_contour_ymid}
\end{figure*}

For a qualitative visualization of the flow inside a tree canopy, we first present the mean streamwise wind velocity and the turbulent kinetic energy, TKE $=\displaystyle  0.5 \left( \overline{u^\prime u^\prime} + \overline{v^\prime v^\prime} + \overline{w^\prime w^\prime}  \right)$,  contours at the mid-span vertical plane for typical representative cases at $\ell_d = 3.33$ m; see Fig.~\ref{fig:u_and_tke_contour_ymid}. The overall qualitative characteristics of the flow through the tree canopy remain the same for all the canopy types.
The presence of the tree slows down the incoming flow as it enters into the canopy. Both flow velocity and TKE decay close to zero towards the leeward side of the canopy that has $L$ ($22.5$ m) sufficiently longer than $\ell_d$ ($3.33$ m) satisfying the condition of continuous canopy $L > 3(1+\beta)\ell_d$; see cases S225, L225 and I225 in Fig.~\ref{fig:u_and_tke_contour_ymid}. 
However, for the canopy W225, the flow velocity does not reach zero at the leeward plane, though it exhibits the decaying trend being consistent with the other cases. In W225, the canopy length ($4.5$ m) is of a similar order of drag length ($3.33$ m) and is within the wind-break regime $L < 3(1+\beta)\ell_d$.

\begin{figure*}[t!]
\centering
\includegraphics[scale=0.245]{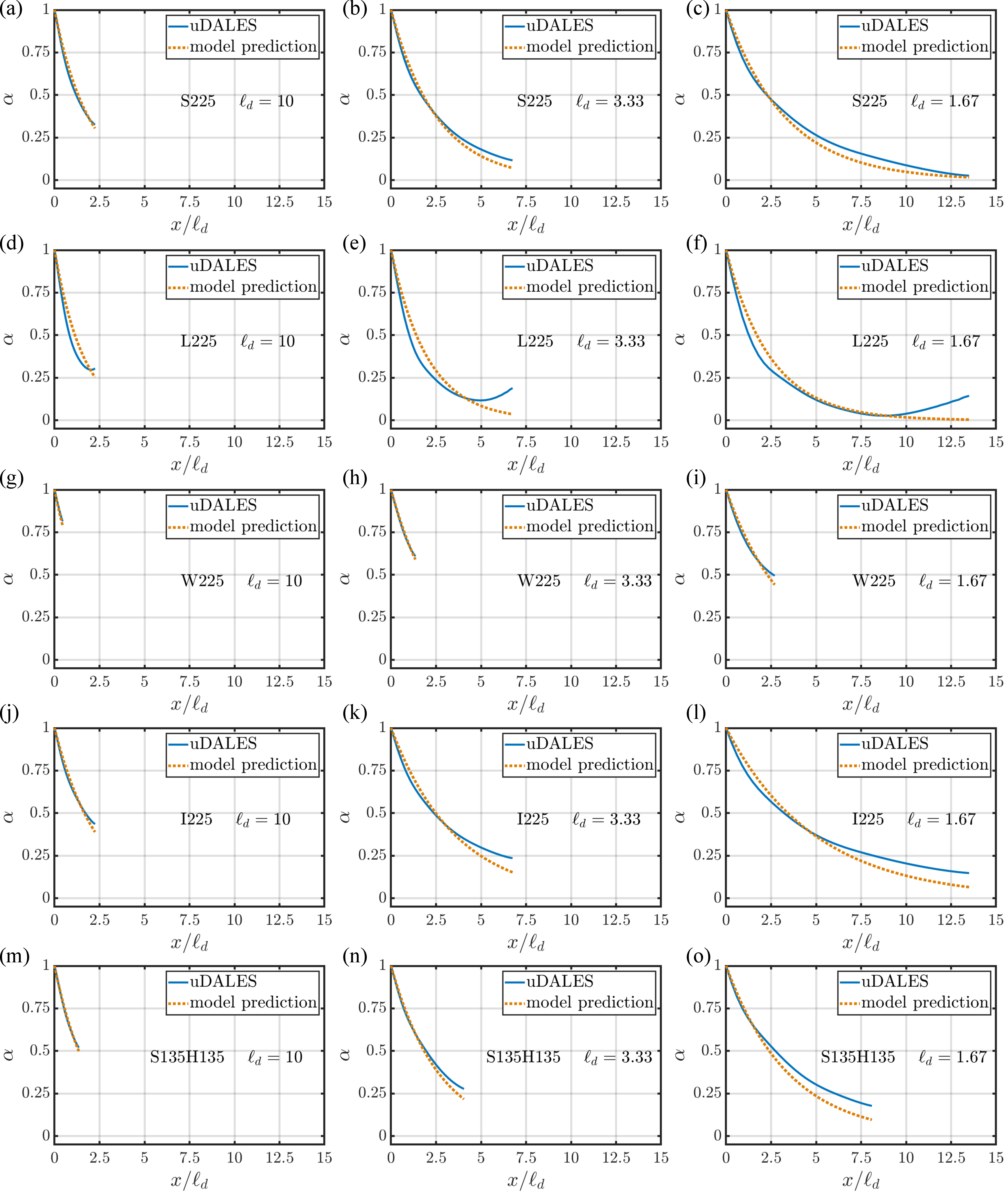}
\caption{Evolution of aerodynamic porosity inside the tree canopy: actual values obtained from simulation versus model prediction for a set of typically chosen cases. Other cases follow the similar trend as well, and not shown here only for the sake of brevity.}
\label{fig:apx_vs_xbielld}
\end{figure*}

To aid quantitative understanding, the variation of aerodynamic porosity along the length of the tree canopy is presented in Fig.~\ref{fig:apx_vs_xbielld} for a few typically chosen cases. Note that the evolution of $\alpha$ computed from the simulation output indeed reflects an exponential decay reasonably well and thus supports Eq.~\eqref{eq:alpha2_6} and the theoretical model proposed in Section~\ref{sec:analytical_model}. 

The coefficient $\beta$ in Eq.~\eqref{eq:alpha2_6} can be estimated by extracting $\alpha(x)$ from the wind velocity data inside the canopy. For each of the $168$ simulations, an appropriate value of $\beta$ has been obtained, that best fits Eq.~\eqref{eq:alpha2_6} to the $\alpha^2$ variation measured from simulation output velocity field. Here, $\alpha^2$ is considered instead of $\alpha$, because $C_d$ has direct impact from $\alpha^2$ (Eq.~\eqref{eq:Cd-alpha}). 

\begin{figure}
\centering
\includegraphics[scale=0.45]{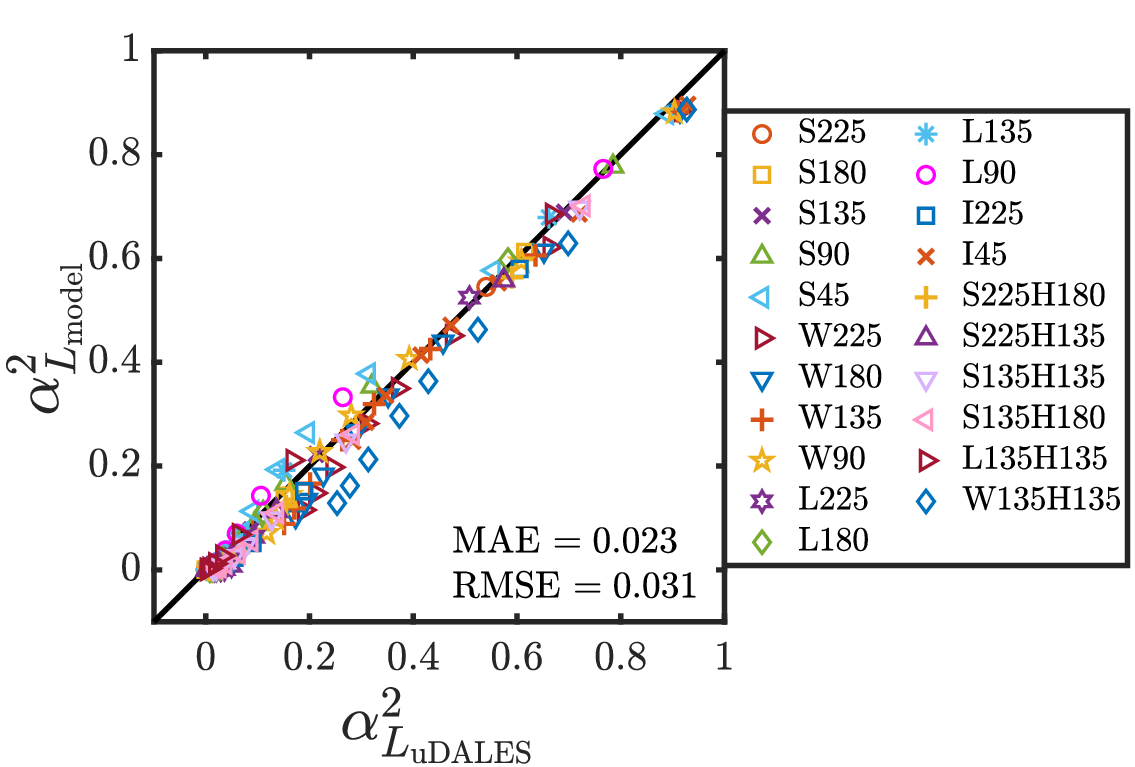}
\caption{Comparison of model prediction and the values obtained from simulation for the square of aerodynamic porosity. MAE: mean absolute error; RMSE: root mean squared error.}
\label{fig:modelprediction_vs_measurement}
\end{figure}

Having obtained $\beta$ for each simulation, a regression analysis was performed. Applying the Buckingham-Pi theorem to the relation $\displaystyle \beta = f\left(L,\,W, \,H, \,\ell_d, \,U_{\infty}, U_0 \right)$, we expect that $\beta = g\left(L/\ell_d, W/\ell_d, H/\ell_d, U_{\infty}/ U_0 \right)$.
Subsequently, a regression analysis was carried out assuming a power-law relation of the identified dimensionless quantities using the MATLAB function\\ \texttt{fminsearch}. Here, only cases for which $\alpha_L>0.3$ were considered \citep[$0.3$ being the critical aerodynamic porosity; ][]{dong2007wind}. The nonlinear fit that minimizes the error norm is given by
\begin{equation}
     \hat{\beta} = 0.27 + 0.55 \left(\frac{WH}{\ell_d^2}\right)^{0.33} \; ,
\label{eq:beta}
\end{equation}
where $\hat{\beta}$ indicates the regression fit for $\beta$. Note that $WH$ corresponds to the frontal area $A_F$. The regression relation in Eq.~\eqref{eq:beta} is a function of the projected frontal area and the drag length only, and not on $L/\ell_d$ or $U_\infty/U_0$. 

In order to test the accuracy of this approximate model, $\hat{\beta}$ was estimated based on the input $L$, $W$ and $\ell_d$ for all the $168$ simulation cases, and then the evolution of $\alpha$ was predicted as per Eq.~\eqref{eq:alpha2_6}. These model predictions are shown in terms of `orange-dotted' lines in Fig.~\ref{fig:apx_vs_xbielld}. A reasonably good agreement is evident. Additionally, the model predicted $\alpha_L^2$ are plotted against its value computed from the simulation output wind velocity data in Fig.~\ref{fig:modelprediction_vs_measurement}, showing mean absolute error (MAE) and root mean square error (RMSE) of only $0.023$ and $0.031$, respectively. $\displaystyle \alpha_{L_{\rm model}}^2$ and $\displaystyle \alpha_{L_{\rm uDALES}}^2$ exhibit reasonably good match, establishing a good confidence on the proposed regression model. Next, we discuss what are implications of the theoretical model proposed in Section~\ref{sec:analytical_model} and the regression model in Eq.~\eqref{eq:beta}, and how these can be put to use in practice.

\section{Implications}\label{sec:implecations}

\subsection{Linking wind-tunnel experiments to CFD simulations} \label{sec:interlink}

The analytical model can be used to estimate $\ell_d$ of model trees used in wind-tunnel experiments. These experiments typically report $C_d$ and $\alpha_L$ (Table~\ref{tab:literature values 2}). As pointed out in Section~\ref{sec:analytical_model}, $\alpha_L$ alone is sufficient to estimate the corresponding drag length. Indeed, rearranging Eq.~\eqref{eq:alpha2_6} and substituting Eq.~\eqref{eq:beta} leads to
\begin{equation}
    \hat{\ell}_d\left[ 1.27 + 0.55 \left(\frac{WH}{\hat{\ell}_d^2}\right)^{0.33} \right] = - \frac{2L}{\ln{\left(\alpha_L^2\right) }}
    \,,
    \label{eq: elld model}
\end{equation}
which is an implicit equation in $\hat{\ell}_d$ that can be solved using a root finding method. The measurement of bulk drag coefficient is redundant here. The representative values for $L$, $W$ and $H$, and $\alpha_L$ can be measured for a model tree in a wind-tunnel experiment and then can be used to obtain $\hat{\ell}_d$ using Eq.~\eqref{eq: elld model}. This estimated value of $\ell_d$ can then be used as input to model the tree in a corresponding CFD simulation to replicate the same drag behaviour. We calculated $\hat{\ell}_d$ for a series of wind-tunnel tree models from literature and summarise them in Table~\ref{tab:literature values 2}. $\hat{\ell}_d$ for different scaled wind-tunnel tree models ranges between $0.003$-$0.6$ m approximately. The present study recommends using $\ell_d$ estimated from Eq.~\eqref{eq: elld model} to model a tree in CFD simulations as this equation is obtained based on the wind velocity (a direct flow-field variable) inside the tree canopy, instead of estimating $\ell_d$ from an average quantity like $C_d$. Note that for lower aerodynamic porosity ($\alpha_L < 0.3$) caused either by long canopy length or due to higher drag length, $\ln \left( \alpha_L^2 \right)$ becomes extremely sensitive to the value of $\alpha_L$, which may lead to erroneous prediction of $\ell_d$ using Eq.~\eqref{eq: elld model}, and the model should be used with caution.

\subsection{Appropriate scaling for model trees in wind tunnel}\label{sec:wt scaling}

Having inferred $\ell_d$ from the wind-tunnel experiments, it becomes possible to argue what the appropriate scaling should be for trees in wind tunnels. To this extent, we non-dimensionalise the Navier-Stokes equations using a characteristic building height $H_b$ and velocity $U_\infty$, which results in
\begin{equation}
\frac{D \vec u}{Dt} = - \nabla p +  \frac{1}{Re} \nabla^2 \vec u - Dr \,\lvert\vec u\rvert \, \vec u + \vec f \; ,
\end{equation}
where $Re = U_{\infty}H_b/\nu$ is the Reynolds number ($\nu$ is the kinematic viscosity), and $Dr$ is the vegetation drag number defined as
\begin{equation}
Dr = \frac{H_b}{\ell_d} \; .
\end{equation}
Wind tunnels often use geometric scaling, \textit{i.e.}, all geometrical dimensions are scaled down by a factor $s$. Ideally all dimensionless quantities, $Re$ and $Dr$, in this case would kept identical to achieve full similarity. However, it is typically not possible to achieve the same values of $Re$ in wind tunnels as in reality, but this is not of much concern as $Re$ does not play a crucial role in the urban settings provided it is larger than $10^3$ based on the building height~\citep{shu2020dimensional}. 
In order to achieve similarity for the vegetative drag, $Dr$ should have the same value at both model scale (MS) and full scale (FS):
In order to perform the wind-tunnel measurements under the same conditions as in reality, the drag length should scale by the factor $s$ as well in order to maintain the same value of $Dr$, \textit{{i.e.}},
\begin{equation}
    \left. \frac{H_b}{\ell_d} \right|_{\rm MS} = \left. \frac{H_b}{\ell_d} \right|_{\rm FS}
    \quad \Leftrightarrow \quad
    \frac{\ell_{d;{\rm MS}}}{\ell_{d;{\rm FS}}}  = \frac{H_{b;{\rm MS}}}{H_{b;{\rm FS}}} = \frac{1}{s} \; ,
\end{equation}
and thus similarity requires that  $\ell_{d;{\rm MS}} = \ell_{d;{\rm FS}} / s$.

The typical scaling factor $s$ in wind tunnels is $100-400$~\citep{wang1996scale,gromke2016influence,gromke2018wind}. From Table~\ref{tab:literature values 1} and Fig.~\ref{fig:elld_box}, the median of full-scale $\ell_{d;{\rm FS}}$ of trees is $21$ m with interquartile range of $10$ to $34$ m. Based on the arguments above, for a wind tunnel with a scaling factor of $100$, $\ell_{d;{\rm MS}}$ needs to be within the interquartile range of $0.1-0.34$ m with median of $0.21$ m. For a scaling factor of $400$, $\ell_{d;{\rm MS}}$ needs to be within the interquartile range of $0.025-0.085$ m with median $0.05$ m. 
From the estimated drag lengths of scaled model trees in Table~\ref{tab:literature values 2}, the median of $\ell_{d;{\rm MS}}$ is approximately $0.05$ m with an interquartile range of $0.03-0.15$ m.
This confirms that trees used in the wind-tunnel experiments have drag length overall falling within the range mentioned above, albeit being on the lower side in comparison to real trees for $s=100$, suggesting the effects of vegetation on drag may be overestimated by wind-tunnel models.

The argument for wind-tunnel scaling put forward here is consistent with that of \citet{gromke2016influence, gromke2018wind}, as demonstrated below. They use foam to represent tree canopies and infer its properties by measuring the pressure loss across a foam model of thickness $L$ that spans the entire wind-tunnel cross-section. They determine a pressure loss coefficient $\lambda$ defined as \citep{gromke2016influence, gromke2018wind, buccolieri2018review}
\begin{equation}
    \label{eq:lambda}
    \lambda = \frac{2 \Delta p}{U^2 L}
\end{equation}
where $\Delta p = p_{ww} - p_{lw}$ is the kinematic pressure difference between the windward and leeward sides of the foam. Note that $\lambda$ has unit m$^{-1}$.

As the foam covers the entire cross-section, the mean velocity $U$ will remain constant and the momentum balance in Eq.~\eqref{eq:momu_tree_ode_6} simply becomes
\begin{equation}
  \frac{1}{c}\frac{\Delta p}{L} = \frac{1}{\ell_d} U^2.
\end{equation}
Rearranging and substituting Eq.~\eqref{eq:lambda} yields
\begin{equation}
  \ell_d = \frac{c U^2 L}{\Delta p} = 2c\lambda^{-1},
\end{equation}
which shows that the drag length $\ell_d$ is simply the inverse of the pressure loss coefficient $\lambda$ up to a constant.

Covering the entire cross-section with porous material is a smart way to avoid three-dimensional flow effects that normal canopies will create otherwise. It offers a straightforward manner to determine $\ell_d$ without having to involve the model developed in this paper.
However, if one wants to determine $\ell_d$ of trees including trunks it will be impossible to cover the entire wind tunnel uniformly and the method developed in this paper will be preferable.

\subsection{Wind engineering}
\label{sec:wind engineering} 

Wind engineering practitioners have increasingly adopted a numerical (CFD) approach to model the pedestrian level wind environment and provide feedback on comfort and safety conditions for the public. These studies are part of several councils’ strategies to improve the liveability of cities, as they become denser and taller due to a rapid increase in urbanisation. A notable case in the UK is the City of London, that recently defined specific Wind Microclimate Guidelines for new developments in their borough~\citep{COLguideline2019}.

One of the key aspects of wind microclimate studies is the development and assessment of wind mitigation measures to resolve or dissipate any excessive windiness in and around the area of interest. Practitioners rely on CFD models, often steady-state, to quantify the effectiveness of these mitigations in comparison with the baseline (unmitigated) situation. The mitigation measures include small elements, however, their geometric and aerodynamic properties are hard to reproduce in urban CFD models and may lead to large uncertainties in the results. This is particularly true for porous elements such as trees and vegetation. Among the many sources of uncertainties in commercial CFD models of trees, some are related to the choice of appropriate aerodynamic parameters, on their use in the numerical model (particularly in the turbulence terms), and on the sensitivity to the grid resolution. 

Due to the lack of specific guidelines in the market, the choice of appropriate parameters to describe the aerodynamic properties of vegetation is up to the modeller. Table~\ref{tab:literature values 1} illustrates the large variability of $a$ and $C_d^V$ found in literature, which inevitably leads to inconsistencies in simulation results from different providers. This paper introduces a way to combine these variables into the drag length $\ell_d$, a physically meaningful lengthscale that describes the spatial extent to which the tree exerts influence on the wind. By consolidating the input values into one parameter, the drag length has the potential to reduce the spread in the results. However, further work is needed to provide an extensive dataset and guidelines for practitioners to enable a wider understanding and use of $\ell_d$ in different situations.

One pertinent problem is that $a$ formally only provides information about the leaves. For wind engineering applications, the trunks are also of importance, particularly since winter situations with leave-less trees are typically used to quantify the effect of vegetation on wind microclimate. Here, the drag length $\ell_d$ is highly beneficial, since it directly quantifies the drag, and the parameters $C_d^V$ and $a$ simply need to be set such that the appropriate value of $\ell_d$ is obtained. For winter-scenarios, this might involve taking a normal value for $C_D^V$ (say $0.2$) and then calculating the leaf-area density as $(C_d^V \ell_d)^{-1}$. In CFD simulations of summer scenarios that include tree's evaporative processes, it is paramount the correct value of $a$ is used, and thus $C_d^V$ can be set as $(a \ell_d)^{-1}$ to capture both drag and evaporation appropriately.

Furthermore, the drag length $\ell_d$ is likely dependent on the wind speed. Indeed, it is well known that trees undergo foliage reconfiguration at high wind speeds~\citep{bekkers2022drag}, which creates a strong dependence of $C_d^V$ on wind velocity \citep{manickathan2018comparative}. 
For this reason some authors propose to use a streamlining coefficient to be combined with $C_d^V$, depending on structural properties of the tree crown~\citep{rudnicki2004wind,vollsinger2005wind,manickathan2018comparative}.

A considerable source of uncertainty in commercial tree simulations is the strong dependency of the results on the grid resolution in the topology representing a tree in the CFD model. The concept of drag length $\ell_d$ is linked to how rapidly the flow is retarded inside the tree canopy. This explicit connection between the tree aerodynamic properties and their effects on the flow can provide useful insights to the minimum grid resolution that is needed to resolve the flow in the area surrounding the tree, which would be some fraction of $\ell_d$. This could be very valuable for the industry and guide the definition of specific guidelines for tree modelling in CFD. We thus recommend further work to be carried out that explores the grid requirements due to vegetation with simulation codes used by practitioners.

\section{Conclusions}
\label{sec:con}

This study developed an analytical model to evaluate drag characteristics of wind-breaks. The model identifies a critical drag parameter, namely the drag length $\ell_d = (a C_d^V)^{-1}$. Here, $a$ is the leaf-area density and $C_d^V$ is a volumetric drag coefficient, not to be confused with the classical (bulk) drag coefficient $C_d$ that is often determined based on projected frontal area in wind-tunnel experiments of vegetation.
A detailed study of the literature, summarised in Table~\ref{tab:literature values 1}, demonstrates that the median value of $\ell_d$ observed in field experiments is $21$ m. The median value of $\ell_d$ for low vegetation (\textit{e.g.}\ crops) is $0.7$ m. For tree canopy and low vegetation, there is a substantial spread in the data.

The analytical model clearly shows that the bulk drag coefficient $C_d$ is linked to the aerodynamic porosity $\alpha$ as $C_d \sim (1-\alpha^2)$, providing an explanation for the strong correlation between these parameters observed in wind-tunnel studies.
The extensive parametric LES investigation conducted in this work allowed to obtain a closed form of the analytical model, which permits the direct translation between wind-tunnel experiments (that tend to determine $C_d$ and $\alpha$) and CFD simulations (which require $a$ and $C_d^V$). This makes it possible to understand what values for $a$ and $C_d^V$ are required to perform simulations of wind-tunnel experiments, and vice versa. 
The calculation of $\ell_d$ for a substantial number of wind-tunnel experiments was performed in Table~\ref{tab:literature values 2}.  
Median $\ell_d$ is $0.05$ m with an interquartile range $0.03 - 0.15$ m for the scaled tree models in wind tunnels.

The identification of the drag length $\ell_d$ provides clarity on what scaling needs to be used for trees in wind-tunnel experiments. Through the non-dimensionalisation of the Navier-Stokes equations, it is possible to derive a dimensionless vegetation drag number $Dr = \ell_d/H_b$, with $H_b$ a characteristic length of the problem. 
Geometric scaling would therefore need to be applied to $\ell_d$ as well in order to achieve the same dynamic conditions as at full scale. A comparison with the work of \cite{gromke2016influence, gromke2018wind} showed full consistency with the scaling proposed in that work.

In the context of wind engineering, the introduction of a new vegetation metric would bring about significant benefits to the current status quo. 
Indeed, reducing the uncertainty in tree modelling is essential if we wish to include trees in wind mitigation schemes with confidence. 
Future efforts should focus on expanding the dataset of drag length values across various tree species and sizes, as well as accounting for seasonal variations in deciduous trees. By doing so, wind engineers will be better equipped to integrate trees and vegetation into their mitigation designs, enhancing not only wind comfort but also delivering valuable environmental co-benefits.

\section*{Author contributions}

D.M.\ and M.v.R.\ conceptualized the research idea. D.M.\ contributed to the analytical model development and carried out the literature survey, numerical simulations, data analysis, figure production (Sects.\ 
\ref{sec:analytical_model}, \ref{sec:simulations}, \ref{sec:results}, \ref{sec:interlink}, \ref{app:drag-coefficient}). G.V., R.R.\ and N.G.\ contributed to the discussion in the context of wind engineering (Sect. \ref{sec:wind engineering}). M.v.R.\ acquired funding, supervised the project as a whole and conceived of the analytical model (Sect. \ref{sec:analytical_model}) and its implications (Sects.\  \ref{sec:interlink}, \ref{sec:wt scaling}, \ref{sec:wind engineering}). All the authors contributed in preparing and reviewing the manuscript.

\section*{Acknowledgements}

D.M.\ acknowledges financial support through the Turbulence at the Exascale project (EP/W026686/1) which is part of the ExCALIBUR HPC programme funded by EPSRC. M.v.R.\ and D.M.\ appreciate useful discussions with Prof.\ Pietro Salizzoni and Dr. Sofia Fellini about the implications of this work for wind-tunnel experiments. This work was inspired by the guide we are developing on CFD of vegetated urban areas as part of the UK Urban Environment Quality (UKUEQ) initiative which is supported by the Wind Engineering Society and CIBSE.

\appendix

\section{Drag coefficients in literature}\label{app:drag-coefficient}

The typical values of the volumetric drag coefficient and leaf-area density for various types of vegetation canopies are summarised in Table~\ref{tab:literature values 1}. The table includes measurements based on field studies and wind-tunnel experiments of different vegetation species, as well as values used in numerical simulations for modelling vegetation canopies in urban areas. The corresponding drag length for each of these cases can be directly computed as $\displaystyle \left(aC_d^V\right)^{-1}$; see Table~\ref{tab:literature values 1}.
In Table~\ref{tab:literature values 2}, we infer the drag length of the model trees used in different wind-tunnel studies. 
Based on the tree dimension and aerodynamic porosity mentioned in these studies, the corresponding values of $\hat{\ell}_d$ are estimated by solving Eq.~\eqref{eq: elld model} numerically using the MATLAB function \texttt{solve}. 
The bulk drag coefficient values are redundant in this context. The $\hat{\ell}_d$ values can be directly used to replicate these wind-tunnel studies in CFD simulations.

\begin{table*}[t!]
\resizebox{0.75\textwidth}{!}{%
\begin{threeparttable}
\small
\caption{Typical values of drag coefficient and leaf-area density reported/used in literature for different vegetation canopies.}
\label{tab:literature values 1}
\begin{center}
\begin{tabular}{p{4cm}p{4.5cm}p{1.3cm}p{2.5cm}p{2.5cm}}
	\textbf{Reference} & \textbf{Type} &  $C_d^{V}$ [-] &  $a$ [m$^{-1}$] &  $\ell_d$ [m] \\
\hline
\hline
    \multicolumn{2}{l}{\textbf{Field experiments of tree canopies}} &  &  &  \\

        \citet{baldocchi1988spectral} & deciduous & $0.15$\tnote{b} & $0.28$\tnote{a} & $23.61$  \\

        \citet{amiro1990drag} & aspen & $0.14$\tnote{a} & $0.43$\tnote{a} & $16.67$  \\

        & pine & $0.171$\tnote{a} & $0.166$\tnote{a} & $35.06$  \\

        & spruce & $0.13$\tnote{a} & $1.59$\tnote{a} & $4.85$  \\

        \citet{gardiner1994wind} & spruce & $0.2$\tnote{b} & $0.57$\tnote{a} & $8.77$  \\

        \citet{katul2004one} & spruce & $0.2$ & $1.0$ & $5.0$  \\

        & aspen & $0.2$ & $0.4$ & $12.5$  \\

        & jack pine & $0.2$ & $0.133$ & $37.5$  \\

        & scots pine & $0.2$ & $0.13$ & $38.5$  \\

        & loblolly pine & $0.2$ & $0.237$ & $21.05$  \\

        & hardwood forest & $0.15$ & $0.227$ & $29.33$  \\

\hline
    \multicolumn{2}{l}{\textbf{Field/wind-tunel experiments of low vegetation}} &  &  &  \\

    \citet{shaw1974measurements,massman1987comparative} & corn canopy & $0.17$ & $1.14$\tnote{a} & $5.15$  \\

    \citet{wilson1982statistics} & corn canopy & $0.17$ & $3.03$\tnote{a} & $1.94$  \\

    \citet{katul2004one} & rice canopy & $0.3$ & $4.31$ & $0.77$  \\

    & corn canopy & $0.3$ & $1.32$ & $2.53$  \\
        
    \citet{molina2006wind}\tnote{c} & tomato & $0.26$ & $5.6$, $8.2$, $11.7$ & $0.69$, $0.47$, $0.33$  \\

        & sweet pepper & $0.23$ & $5.8$, $10.6$, $12.6$ & $0.75$, $0.41$, $0.34$  \\

        & aubergine & $0.23$ & $3.7$, $6.7$, $11.6$ & $1.17$, $0.65$ $0.37$  \\

        & bean & $0.22$ & $3.6$, $5.8$, $6.8$ & $1.26$, $0.78$ $0.67$  \\
        
\hline
    \multicolumn{2}{l}{\textbf{Numerical simulations of vegetation canopies}} &  &  &  \\

        \citet{li1985first} & pine forest, corn canopy & $0.165$ & $0.71$\tnote{a}, $1.06$\tnote{a} & $8.54$, $5.72$ \\
        
        \citet{shaw1992large} & deciduous forest & $0.15$ & $1.99$\tnote{a}, $5.04$\tnote{a} & $3.34$, $1.32$ \\

        \citet{svensson1990two} & corn canopy, orchard & $0.3$ & $0.5$, $2.1$ & $6.67$, $1.59$ \\
    
        \citet{liang2006improved} & trees in an urban area & $0.2$ & $0.4$ - $0.75$ & $12.5$ - $6.67$ \\

        \citet{belcher2008flows} & open woodland to dense spruce plantation & $0.25$ & $0.1$ - $1.0$ & $40.0$ - $4.0$ \\
        
        \citet{amorim2013cfd} & trees in an urban area & $0.2$ & $1.0$ & $5.0$  \\

        \citet{kenjerevs2013modelling} & trees in an urban area & $0.1$ & $1.0$, $3.0$ & $10.0$, $3.33$  \\
        \citet{gromke2015influencea,gromke2015influenceb} & trees in an urban area & $0.2$ & $1.0$ & $5.0$  \\
        
        \citet{gromke2015cfd} & trees in an urban area & $0.2$ & $0.55$, $0.75$, $1.5$  & $9.1$, $6.67$, $3.33$  \\
        
        \citet{krayenhoff2015parametrization} & trees in an urban area & $0.2$ & $0.06$ - $0.5$ & $83.33$ - $10$  \\
        
        \citet{vranckx2015impact} & trees in an urban area & $0.15$, $0.33$ & $1.6$, $4.0$ & $4.17$, $1.89$, $0.76$  \\

        \citet{ghasemian2017influence} & trees in an urban area & $0.6$ & $0.17$, $0.42$, $1.0$, $1.25$, $3.33$ & $9.8$, $3.97$, $1.67$, $1.33$, $0.5$  \\
        
        \citet{jeanjean2017air} & trees in an urban area & $0.25$ & $0.0$, $1.06$, $1.6$ & $\infty$, $3.85$, $2.5$  \\
        
        \citet{moradpour2017numerical} & trees in an urban street canyon & $0.2$ & $0.5$ - $2.0$ & $10.0$ - $2.5$  \\
            
        \citet{santiago2017dry} & trees in an urban area & $0.2$ & $0.0625$, $0.125$, $0.25$ & $80.0$, $40.0$, $20.0$  \\

        \citet{yang2017numerical} & trees in subtropical urban park & $0.2$ & $1.0$, $4.0$ & $5.0$, $1.25$  \\
        
        \citet{grylls2021tree,grylls2022trees} & trees in an urban area & $0.2$ & $1.0$ & $5.0$  \\

        \citet{ricci2022impact} & trees in an urban area & $0.3$ & $1.0$ - $3.0$ & $3.33$ - $1.11$  \\
        
        \citet{duan2024modulating} & trees in an urban area & $0.1$, $0.2$ & $2.5$, $1.0$ & $4.0$, $5.0$  \\
        
        \citet{fu2024should} & trees in an urban area & $0.2$ & $2.2$, $1.8$, $1.6$, $1.4$, $1.0$, $0.6$, $0.2$ & $2.27$, $2.78$, $3.12$, $3.57$, $5.0$, $8.33$, $25.0$  \\
\hline
\end{tabular}
\begin{tablenotes}
\footnotesize
    \item[a] average value calculated based on the profile given in the article
    \item[b] taken based on other literature of similar vegetation type
    \item[c] this study was carried out in wind tunnel
\end{tablenotes}
\end{center}
\end{threeparttable}
}
\end{table*}

\begin{table*}[t!]
\begin{threeparttable}[t!]
\small
\caption{Estimated drag length for wind-tunnel model trees from literature. }
\label{tab:literature values 2}
\begin{center}

\begin{tabular}{p{3.5cm}p{2.2cm}p{1.4cm}p{0.8cm}p{1.2cm}p{0.8cm}p{0.9cm}p{0.8cm}p{0.8cm}}
	\textbf{Reference} & \textbf{Description} & $C_d$ [-] & $\alpha_L$ [-] & $U_{\infty} [m/s]$ & $L$ [m] & $W$ [m] & $H$ [m] & $\hat{\ell_d}$ [m] \\
\hline
\hline
    \citet{grant1998direct} & \multicolumn{2}{l}{artificial Scots pine Christmas tree}  &  &  \\
        & - least porous & $1.2$\tnote{a} & $0.51$\tnote{b} & -- & $0.3$\tnote{c} & $0.3$ & $1.45$ & $0.173$\\
        & - medium porous & $1.07$\tnote{a} & $0.63$\tnote{b} & -- & $0.3$\tnote{c} & $0.3$ & $1.45$ & $0.299$\\
        & - most porous & $0.865$\tnote{a} & $0.69$\tnote{b} & -- & $0.3$\tnote{c} & $0.3$ & $1.45$ & $0.402$ \\
    
    \citet{guan2003wind} & model no. 1 & $1.06$\tnote{a} & $0.133$ & $1.6-5.3$ & $0.14$ & $0.5$ & $0.1$ & $0.015$ \\
        & model no. 2 & $0.94$\tnote{a} & $0.303$ & $1.6-5.3$ & $0.14$ & $0.5$ & $0.1$ & $0.038$ \\
        & model no. 3 & $0.84$\tnote{a} & $0.401$ & $1.6-5.3$ & $0.1$ & $0.5$ & $0.1$ & $0.034$ \\
        & model no. 4 & $0.81$\tnote{a} & $0.450$ & $1.6-5.3$ & $0.1$ & $0.5$ & $0.1$ & $0.043$ \\
        & model no. 5 & $0.74$\tnote{a} & $0.503$ & $1.6-5.3$ & $0.1$ & $0.5$ & $0.1$ & $0.054$ \\
        & model no. 6 & $0.67$\tnote{a} & $0.605$ & $1.6-5.3$ & $0.1$ & $0.5$ & $0.1$ & $0.086$ \\
        & model no. 7 & $0.6$\tnote{a} & $0.685$ & $1.6-5.3$ & $0.1$ & $0.5$ & $0.1$ & $0.128$ \\

    \citet{bitog2011wind}\tnote{e} & \emph{Pinus thunbergii} &  &  &  &  &  &  &  \\
    
        & -- one tree & $0.55$ & $0.91$ & $2.0-8.0$ & $1.0$ & $1.0$ & $1.7$ & $7.34$ \\
            
		& -- two tree & $0.82$ & $0.69$ & $2.0-8.0$ & $1.0$ & $1.0$ & $1.7$ & $1.53$ \\
      
         & -- three tree & $1.08$ & $0.42$ & $2.0-8.0$ & $1.0$ & $1.0$ & $1.7$ & $0.5$ \\

    \citet{lee2014shelter} & \emph{Abides concolor} &  &  &  &  &  &  &  \\
    
        & -- control & -- & $0.2985$ & $5.0$ & $0.11$\tnote{c} & $0.11$ & $0.19$ & $0.033$ \\
     
        & -- rotated & -- & $0.344$ & $5.0$ & $0.11$\tnote{c} & $0.11$ & $0.19$ & $0.04$ \\
     
        & -- no leaf & -- & $0.8782$ & $5.0$ & $0.11$\tnote{c} & $0.11$ & $0.19$ & $0.568$ \\

    \citet{manickathan2018comparative} & model tree 1 & $0.58$ & $0.102$\tnote{b} & $3.0-20.0$ & $0.04$\tnote{c} & $0.04$\tnote{d} & $0.12$ & $0.003$ \\

        & model tree 2 & $0.68$ & $0.219$\tnote{b} & $3.0-20.0$ & $0.082$\tnote{c} & $0.082$\tnote{d} & $0.10$ & $0.019$ \\

        & model tree 3 & $0.70-0.75$ & $0.448$\tnote{b} & $3.0-20.0$ & $0.114$\tnote{c} & $0.114$\tnote{d} & $0.21$ & $0.062$  \\

        & \emph{Chamaecyparis pisifera} & $0.72-0.87$ & $0.381$\tnote{b} & $3.0-20.0$ & $0.119$\tnote{c} & $0.119$\tnote{d} & $0.25$ & $0.048$ \\

        & \emph{Ilex crenata} & $0.87$ & $0.382$\tnote{b} & $3.0-20.0$ & $0.128$\tnote{c} & $0.128$\tnote{d} & $0.32$ & $0.05$ \\

        & \emph{Ilex crenata} -- defoliated & $0.31-0.33$ & $0.871$\tnote{b} & $3.0-20.0$ & $0.128$\tnote{c} & $0.128$\tnote{d} & $0.32$ & $0.606$ \\

    \citet{fellini2022high} & single model tree & $1.07$ & $0.3$ & $4.0-24.0$ & $0.045$ & $0.045$ & $0.065$ & $0.014$ \\

\hline
\end{tabular}

\begin{tablenotes}
\footnotesize
    \item[a] $C_d$ was calculated based on the streamwise velocity at canopy crown height $\langle \overline{u} \rangle_{y}\rvert_{z=H}$, instead of $U_{\infty}$
    \item[b] $\alpha_L$ was calculated from optical porosity ($\alpha_O$) at quiescent condition as $\alpha_L = \alpha_O^{0.4}$~\citep{guan2003wind}
    \item[c] $L$ was not mentioned in the particular literature, hence takes same as $W$
    \item[d] $W$ was not mentioned in the particular literature, hence calculated in a way that $W \times H$ matches the mentioned projected frontal area.
    \item[e] this study considered full-scale tree models
\end{tablenotes}

\end{center}
\end{threeparttable}
\end{table*}

\bibliographystyle{elsarticle-harv} 
\bibliography{cas-refs}

\begin{thebibliography}{70}
\expandafter\ifx\csname natexlab\endcsname\relax\def\natexlab#1{#1}\fi
\providecommand{\url}[1]{\texttt{#1}}
\providecommand{\href}[2]{#2}
\providecommand{\path}[1]{#1}
\providecommand{\DOIprefix}{doi:}
\providecommand{\ArXivprefix}{arXiv:}
\providecommand{\URLprefix}{URL: }
\providecommand{\Pubmedprefix}{pmid:}
\providecommand{\doi}[1]{\href{http://dx.doi.org/#1}{\path{#1}}}
\providecommand{\Pubmed}[1]{\href{pmid:#1}{\path{#1}}}
\providecommand{\bibinfo}[2]{#2}
\ifx\xfnm\relax \def\xfnm[#1]{\unskip,\space#1}\fi
\bibitem[{Amiro(1990)}]{amiro1990drag}
\bibinfo{author}{Amiro, B.}, \bibinfo{year}{1990}.
\newblock \bibinfo{title}{Drag coefficients and turbulence spectra within three boreal forest canopies}.
\newblock \bibinfo{journal}{Boundary-Layer Meteorology} \bibinfo{volume}{52}, \bibinfo{pages}{227--246}.
\bibitem[{Amorim et~al.(2013)Amorim, Rodrigues, Tavares, Valente and Borrego}]{amorim2013cfd}
\bibinfo{author}{Amorim, J.}, \bibinfo{author}{Rodrigues, V.}, \bibinfo{author}{Tavares, R.}, \bibinfo{author}{Valente, J.}, \bibinfo{author}{Borrego, C.}, \bibinfo{year}{2013}.
\newblock \bibinfo{title}{Cfd modelling of the aerodynamic effect of trees on urban air pollution dispersion}.
\newblock \bibinfo{journal}{Science of the Total Environment} \bibinfo{volume}{461}, \bibinfo{pages}{541--551}.
\bibitem[{Baldocchi and Meyers(1988)}]{baldocchi1988spectral}
\bibinfo{author}{Baldocchi, D.D.}, \bibinfo{author}{Meyers, T.P.}, \bibinfo{year}{1988}.
\newblock \bibinfo{title}{A spectral and lag-correlation analysis of turbulence in a deciduous forest canopy}.
\newblock \bibinfo{journal}{Boundary-Layer Meteorology} \bibinfo{volume}{45}, \bibinfo{pages}{31--58}.
\bibitem[{Banerjee et~al.(2013)Banerjee, Katul, Fontan, Poggi and Kumar}]{banerjee2013mean}
\bibinfo{author}{Banerjee, T.}, \bibinfo{author}{Katul, G.}, \bibinfo{author}{Fontan, S.}, \bibinfo{author}{Poggi, D.}, \bibinfo{author}{Kumar, M.}, \bibinfo{year}{2013}.
\newblock \bibinfo{title}{Mean flow near edges and within cavities situated inside dense canopies}.
\newblock \bibinfo{journal}{Boundary-layer meteorology} \bibinfo{volume}{149}, \bibinfo{pages}{19--41}.
\bibitem[{Bekkers et~al.(2022)Bekkers, Angelou and Dellwik}]{bekkers2022drag}
\bibinfo{author}{Bekkers, C.C.}, \bibinfo{author}{Angelou, N.}, \bibinfo{author}{Dellwik, E.}, \bibinfo{year}{2022}.
\newblock \bibinfo{title}{Drag coefficient and frontal area of a solitary mature tree}.
\newblock \bibinfo{journal}{Journal of Wind Engineering and Industrial Aerodynamics} \bibinfo{volume}{220}, \bibinfo{pages}{104854}.
\bibitem[{Belcher et~al.(2008)Belcher, Finnigan and Harman}]{belcher2008flows}
\bibinfo{author}{Belcher, S.}, \bibinfo{author}{Finnigan, J.}, \bibinfo{author}{Harman, I.}, \bibinfo{year}{2008}.
\newblock \bibinfo{title}{Flows through forest canopies in complex terrain}.
\newblock \bibinfo{journal}{Ecological Applications} \bibinfo{volume}{18}, \bibinfo{pages}{1436--1453}.
\bibitem[{Bitog et~al.(2011)Bitog, Lee, Hwang, Shin, Hong, Seo, Mostafa and Pang}]{bitog2011wind}
\bibinfo{author}{Bitog, J.}, \bibinfo{author}{Lee, I.B.}, \bibinfo{author}{Hwang, H.S.}, \bibinfo{author}{Shin, M.H.}, \bibinfo{author}{Hong, S.W.}, \bibinfo{author}{Seo, I.H.}, \bibinfo{author}{Mostafa, E.}, \bibinfo{author}{Pang, Z.}, \bibinfo{year}{2011}.
\newblock \bibinfo{title}{A wind tunnel study on aerodynamic porosity and windbreak drag}.
\newblock \bibinfo{journal}{Forest Science and technology} \bibinfo{volume}{7}, \bibinfo{pages}{8--16}.
\bibitem[{Bozovic et~al.(2017)Bozovic, Maksimovic, Mijic, Smith, Suter and Van~Reeuwijk}]{bozovic2017blue}
\bibinfo{author}{Bozovic, R.}, \bibinfo{author}{Maksimovic, C.}, \bibinfo{author}{Mijic, A.}, \bibinfo{author}{Smith, K.}, \bibinfo{author}{Suter, I.}, \bibinfo{author}{Van~Reeuwijk, M.}, \bibinfo{year}{2017}.
\newblock \bibinfo{title}{Blue green solutions, a systems approach to sustainable, resilient and cost-efficient urban development}.
\newblock \bibinfo{journal}{Technical Report, Climate-KIC, London: Imperial College, UK} .
\bibitem[{Buccolieri et~al.(2018)Buccolieri, Santiago, Rivas and Sanchez}]{buccolieri2018review}
\bibinfo{author}{Buccolieri, R.}, \bibinfo{author}{Santiago, J.L.}, \bibinfo{author}{Rivas, E.}, \bibinfo{author}{Sanchez, B.}, \bibinfo{year}{2018}.
\newblock \bibinfo{title}{Review on urban tree modelling in cfd simulations: Aerodynamic, deposition and thermal effects}.
\newblock \bibinfo{journal}{Urban Forestry \& Urban Greening} \bibinfo{volume}{31}, \bibinfo{pages}{212--220}.
\bibitem[{Chen et~al.(2021)Chen, Yang, Chen, Lam, Hang, Wang, Liu and Ling}]{chen2021integrated}
\bibinfo{author}{Chen, T.}, \bibinfo{author}{Yang, H.}, \bibinfo{author}{Chen, G.}, \bibinfo{author}{Lam, C.K.C.}, \bibinfo{author}{Hang, J.}, \bibinfo{author}{Wang, X.}, \bibinfo{author}{Liu, Y.}, \bibinfo{author}{Ling, H.}, \bibinfo{year}{2021}.
\newblock \bibinfo{title}{Integrated impacts of tree planting and aspect ratios on thermal environment in street canyons by scaled outdoor experiments}.
\newblock \bibinfo{journal}{Science of The Total Environment} \bibinfo{volume}{764}, \bibinfo{pages}{142920}.
\bibitem[{{City of London Corporation}(2019)}]{COLguideline2019}
\bibinfo{author}{{City of London Corporation}}, \bibinfo{year}{2019}.
\newblock \bibinfo{title}{Wind microclimate guidelines for developments in the city of {London}}.
\newblock \bibinfo{howpublished}{\url{https://www.cityoflondon.gov.uk/assets/Services-Environment/wind-microclimate-guidelines.pdf}}.
\newblock \bibinfo{note}{Accessed: 25-Oct-2024}.
\bibitem[{Cullen(2005)}]{cullen2005trees}
\bibinfo{author}{Cullen, S.}, \bibinfo{year}{2005}.
\newblock \bibinfo{title}{Trees and wind: A practical consideration of the drag equation velocity exponent for urban tree risk management}.
\newblock \bibinfo{journal}{Arboriculture \& Urban Forestry (AUF)} \bibinfo{volume}{31}, \bibinfo{pages}{101--113}.
\bibitem[{De-xin et~al.(2000)De-xin, Ting-yao and Shi-jie}]{de2000wind}
\bibinfo{author}{De-xin, G.}, \bibinfo{author}{Ting-yao, Z.}, \bibinfo{author}{Shi-jie, H.}, \bibinfo{year}{2000}.
\newblock \bibinfo{title}{Wind tunnel experiment of drag of isolated tree models in surface boundary layer}.
\newblock \bibinfo{journal}{Journal of Forestry Research} \bibinfo{volume}{11}, \bibinfo{pages}{156--160}.
\bibitem[{Dong et~al.(2007)Dong, Luo, Qian and Wang}]{dong2007wind}
\bibinfo{author}{Dong, Z.}, \bibinfo{author}{Luo, W.}, \bibinfo{author}{Qian, G.}, \bibinfo{author}{Wang, H.}, \bibinfo{year}{2007}.
\newblock \bibinfo{title}{A wind tunnel simulation of the mean velocity fields behind upright porous fences}.
\newblock \bibinfo{journal}{Agricultural and Forest Meteorology} \bibinfo{volume}{146}, \bibinfo{pages}{82--93}.
\bibitem[{Dong et~al.(2008)Dong, Mu, Luo, Qinan, Lu and Wang}]{dong2008analysis}
\bibinfo{author}{Dong, Z.}, \bibinfo{author}{Mu, Q.}, \bibinfo{author}{Luo, W.}, \bibinfo{author}{Qinan, G.}, \bibinfo{author}{Lu, P.}, \bibinfo{author}{Wang, H.}, \bibinfo{year}{2008}.
\newblock \bibinfo{title}{An analysis of drag force and moment for upright porous wind fences}.
\newblock \bibinfo{journal}{Journal of Geophysical Research: Atmospheres} \bibinfo{volume}{113}.
\bibitem[{Duan et~al.(2024)Duan, Bi, Zhao, Yang and Takemi}]{duan2024modulating}
\bibinfo{author}{Duan, G.}, \bibinfo{author}{Bi, Z.}, \bibinfo{author}{Zhao, L.}, \bibinfo{author}{Yang, T.}, \bibinfo{author}{Takemi, T.}, \bibinfo{year}{2024}.
\newblock \bibinfo{title}{Modulating local winds and turbulence around a single building obstacle with the obstruction of tall vegetation}.
\newblock \bibinfo{journal}{Physics of Fluids} \bibinfo{volume}{36}.
\bibitem[{Fellini et~al.(2022)Fellini, Marro, Del~Ponte, Barulli, Soulhac, Ridolfi and Salizzoni}]{fellini2022high}
\bibinfo{author}{Fellini, S.}, \bibinfo{author}{Marro, M.}, \bibinfo{author}{Del~Ponte, A.V.}, \bibinfo{author}{Barulli, M.}, \bibinfo{author}{Soulhac, L.}, \bibinfo{author}{Ridolfi, L.}, \bibinfo{author}{Salizzoni, P.}, \bibinfo{year}{2022}.
\newblock \bibinfo{title}{High resolution wind-tunnel investigation about the effect of street trees on pollutant concentration and street canyon ventilation}.
\newblock \bibinfo{journal}{Building and Environment} \bibinfo{volume}{226}, \bibinfo{pages}{109763}.
\bibitem[{Finnigan(2000)}]{finnigan2000turbulence}
\bibinfo{author}{Finnigan, J.}, \bibinfo{year}{2000}.
\newblock \bibinfo{title}{Turbulence in plant canopies}.
\newblock \bibinfo{journal}{Annual review of fluid mechanics} \bibinfo{volume}{32}, \bibinfo{pages}{519--571}.
\bibitem[{Finnigan and Brunet(1995)}]{finnigan1995turbulent}
\bibinfo{author}{Finnigan, J.J.}, \bibinfo{author}{Brunet, Y.}, \bibinfo{year}{1995}.
\newblock \bibinfo{title}{Turbulent airflow in forests on flat and hilly terrain}.
\newblock \bibinfo{journal}{Wind and trees} , \bibinfo{pages}{3--40}.
\bibitem[{Fu et~al.(2024)Fu, Padjen and Garcia-Sanchez}]{fu2024should}
\bibinfo{author}{Fu, R.}, \bibinfo{author}{Padjen, I.}, \bibinfo{author}{Garcia-Sanchez, C.}, \bibinfo{year}{2024}.
\newblock \bibinfo{title}{Should we care about the level of detail in trees when running urban microscale simulations?}
\newblock \bibinfo{journal}{Sustainable Cities and Society} \bibinfo{volume}{101}, \bibinfo{pages}{105143}.
\bibitem[{Gardiner(1994)}]{gardiner1994wind}
\bibinfo{author}{Gardiner, B.A.}, \bibinfo{year}{1994}.
\newblock \bibinfo{title}{Wind and wind forces in a plantation spruce forest}.
\newblock \bibinfo{journal}{Boundary-Layer Meteorology} \bibinfo{volume}{67}, \bibinfo{pages}{161--186}.
\bibitem[{Ghasemian et~al.(2017)Ghasemian, Amini and Princevac}]{ghasemian2017influence}
\bibinfo{author}{Ghasemian, M.}, \bibinfo{author}{Amini, S.}, \bibinfo{author}{Princevac, M.}, \bibinfo{year}{2017}.
\newblock \bibinfo{title}{The influence of roadside solid and vegetation barriers on near-road air quality}.
\newblock \bibinfo{journal}{Atmospheric Environment} \bibinfo{volume}{170}, \bibinfo{pages}{108--117}.
\bibitem[{Grant and Nickling(1998)}]{grant1998direct}
\bibinfo{author}{Grant, P.}, \bibinfo{author}{Nickling, W.}, \bibinfo{year}{1998}.
\newblock \bibinfo{title}{Direct field measurement of wind drag on vegetation for application to windbreak design and modelling}.
\newblock \bibinfo{journal}{Land Degradation \& Development} \bibinfo{volume}{9}, \bibinfo{pages}{57--66}.
\bibitem[{Gromke(2018)}]{gromke2018wind}
\bibinfo{author}{Gromke, C.}, \bibinfo{year}{2018}.
\newblock \bibinfo{title}{Wind tunnel model of the forest and its reynolds number sensitivity}.
\newblock \bibinfo{journal}{Journal of Wind Engineering and Industrial Aerodynamics} \bibinfo{volume}{175}, \bibinfo{pages}{53--64}.
\bibitem[{Gromke and Blocken(2015a)}]{gromke2015influencea}
\bibinfo{author}{Gromke, C.}, \bibinfo{author}{Blocken, B.}, \bibinfo{year}{2015}a.
\newblock \bibinfo{title}{Influence of avenue-trees on air quality at the urban neighborhood scale. part i: Quality assurance studies and turbulent schmidt number analysis for rans cfd simulations}.
\newblock \bibinfo{journal}{Environmental Pollution} \bibinfo{volume}{196}, \bibinfo{pages}{214--223}.
\bibitem[{Gromke and Blocken(2015b)}]{gromke2015influenceb}
\bibinfo{author}{Gromke, C.}, \bibinfo{author}{Blocken, B.}, \bibinfo{year}{2015}b.
\newblock \bibinfo{title}{Influence of avenue-trees on air quality at the urban neighborhood scale. part ii: Traffic pollutant concentrations at pedestrian level}.
\newblock \bibinfo{journal}{Environmental Pollution} \bibinfo{volume}{196}, \bibinfo{pages}{176--184}.
\bibitem[{Gromke et~al.(2015)Gromke, Blocken, Janssen, Merema, van Hooff and Timmermans}]{gromke2015cfd}
\bibinfo{author}{Gromke, C.}, \bibinfo{author}{Blocken, B.}, \bibinfo{author}{Janssen, W.}, \bibinfo{author}{Merema, B.}, \bibinfo{author}{van Hooff, T.}, \bibinfo{author}{Timmermans, H.}, \bibinfo{year}{2015}.
\newblock \bibinfo{title}{Cfd analysis of transpirational cooling by vegetation: Case study for specific meteorological conditions during a heat wave in arnhem, netherlands}.
\newblock \bibinfo{journal}{Building and environment} \bibinfo{volume}{83}, \bibinfo{pages}{11--26}.
\bibitem[{Gromke et~al.(2016)Gromke, Jamarkattel and Ruck}]{gromke2016influence}
\bibinfo{author}{Gromke, C.}, \bibinfo{author}{Jamarkattel, N.}, \bibinfo{author}{Ruck, B.}, \bibinfo{year}{2016}.
\newblock \bibinfo{title}{Influence of roadside hedgerows on air quality in urban street canyons}.
\newblock \bibinfo{journal}{Atmospheric environment} \bibinfo{volume}{139}, \bibinfo{pages}{75--86}.
\bibitem[{Grylls and van Reeuwijk(2021)}]{grylls2021tree}
\bibinfo{author}{Grylls, T.}, \bibinfo{author}{van Reeuwijk, M.}, \bibinfo{year}{2021}.
\newblock \bibinfo{title}{Tree model with drag, transpiration, shading and deposition: Identification of cooling regimes and large-eddy simulation}.
\newblock \bibinfo{journal}{Agricultural and Forest Meteorology} \bibinfo{volume}{298}, \bibinfo{pages}{108288}.
\bibitem[{Grylls and van Reeuwijk(2022)}]{grylls2022trees}
\bibinfo{author}{Grylls, T.}, \bibinfo{author}{van Reeuwijk, M.}, \bibinfo{year}{2022}.
\newblock \bibinfo{title}{How trees affect urban air quality: It depends on the source}.
\newblock \bibinfo{journal}{Atmospheric Environment} \bibinfo{volume}{290}, \bibinfo{pages}{119275}.
\bibitem[{Guan et~al.(2003)Guan, Zhang and Zhu}]{guan2003wind}
\bibinfo{author}{Guan, D.}, \bibinfo{author}{Zhang, Y.}, \bibinfo{author}{Zhu, T.}, \bibinfo{year}{2003}.
\newblock \bibinfo{title}{A wind-tunnel study of windbreak drag}.
\newblock \bibinfo{journal}{Agricultural and forest meteorology} \bibinfo{volume}{118}, \bibinfo{pages}{75--84}.
\bibitem[{Hagen and Skidmore(1971)}]{hagen1971windbreak}
\bibinfo{author}{Hagen, L.}, \bibinfo{author}{Skidmore, E.}, \bibinfo{year}{1971}.
\newblock \bibinfo{title}{Windbreak drag as influenced by porosity}.
\newblock \bibinfo{journal}{Transactions of the ASAE} \bibinfo{volume}{14}, \bibinfo{pages}{464--0465}.
\bibitem[{Jeanjean et~al.(2017)Jeanjean, Buccolieri, Eddy, Monks and Leigh}]{jeanjean2017air}
\bibinfo{author}{Jeanjean, A.P.}, \bibinfo{author}{Buccolieri, R.}, \bibinfo{author}{Eddy, J.}, \bibinfo{author}{Monks, P.S.}, \bibinfo{author}{Leigh, R.J.}, \bibinfo{year}{2017}.
\newblock \bibinfo{title}{Air quality affected by trees in real street canyons: The case of marylebone neighbourhood in central london}.
\newblock \bibinfo{journal}{Urban Forestry \& Urban Greening} \bibinfo{volume}{22}, \bibinfo{pages}{41--53}.
\bibitem[{Katul et~al.(2004)Katul, Mahrt, Poggi and Sanz}]{katul2004one}
\bibinfo{author}{Katul, G.G.}, \bibinfo{author}{Mahrt, L.}, \bibinfo{author}{Poggi, D.}, \bibinfo{author}{Sanz, C.}, \bibinfo{year}{2004}.
\newblock \bibinfo{title}{One-and two-equation models for canopy turbulence}.
\newblock \bibinfo{journal}{Boundary-layer meteorology} \bibinfo{volume}{113}, \bibinfo{pages}{81--109}.
\bibitem[{Kenjere{\v{s}} and ter Kuile(2013)}]{kenjerevs2013modelling}
\bibinfo{author}{Kenjere{\v{s}}, S.}, \bibinfo{author}{ter Kuile, B.}, \bibinfo{year}{2013}.
\newblock \bibinfo{title}{Modelling and simulations of turbulent flows in urban areas with vegetation}.
\newblock \bibinfo{journal}{Journal of Wind Engineering and Industrial Aerodynamics} \bibinfo{volume}{123}, \bibinfo{pages}{43--55}.
\bibitem[{Krayenhoff et~al.(2015)Krayenhoff, Santiago, Martilli, Christen and Oke}]{krayenhoff2015parametrization}
\bibinfo{author}{Krayenhoff, E.}, \bibinfo{author}{Santiago, J.L.}, \bibinfo{author}{Martilli, A.}, \bibinfo{author}{Christen, A.}, \bibinfo{author}{Oke, T.}, \bibinfo{year}{2015}.
\newblock \bibinfo{title}{Parametrization of drag and turbulence for urban neighbourhoods with trees}.
\newblock \bibinfo{journal}{Boundary-Layer Meteorology} \bibinfo{volume}{156}, \bibinfo{pages}{157--189}.
\bibitem[{Lee et~al.(2014)Lee, Lee and Lee}]{lee2014shelter}
\bibinfo{author}{Lee, J.P.}, \bibinfo{author}{Lee, E.J.}, \bibinfo{author}{Lee, S.J.}, \bibinfo{year}{2014}.
\newblock \bibinfo{title}{Shelter effect of a fir tree with different porosities}.
\newblock \bibinfo{journal}{Journal of Mechanical Science and Technology} \bibinfo{volume}{28}, \bibinfo{pages}{565--572}.
\bibitem[{Li et~al.(1985)Li, Miller and Lin}]{li1985first}
\bibinfo{author}{Li, Z.}, \bibinfo{author}{Miller, D.}, \bibinfo{author}{Lin, J.}, \bibinfo{year}{1985}.
\newblock \bibinfo{title}{A first-order closure scheme to describe counter-gradient momentum transport in plant canopies}.
\newblock \bibinfo{journal}{Boundary-layer meteorology} \bibinfo{volume}{33}, \bibinfo{pages}{77--83}.
\bibitem[{Liang et~al.(2006)Liang, Xiaofeng, Borong and Yingxin}]{liang2006improved}
\bibinfo{author}{Liang, L.}, \bibinfo{author}{Xiaofeng, L.}, \bibinfo{author}{Borong, L.}, \bibinfo{author}{Yingxin, Z.}, \bibinfo{year}{2006}.
\newblock \bibinfo{title}{Improved k--$\varepsilon$ two-equation turbulence model for canopy flow}.
\newblock \bibinfo{journal}{Atmospheric Environment} \bibinfo{volume}{40}, \bibinfo{pages}{762--770}.
\bibitem[{Lyu et~al.(2020)Lyu, Wang and Mason}]{lyu2020review}
\bibinfo{author}{Lyu, J.}, \bibinfo{author}{Wang, C.M.}, \bibinfo{author}{Mason, M.S.}, \bibinfo{year}{2020}.
\newblock \bibinfo{title}{Review of models for predicting wind characteristics behind windbreaks}.
\newblock \bibinfo{journal}{Journal of Wind Engineering and Industrial Aerodynamics} \bibinfo{volume}{199}, \bibinfo{pages}{104117}.
\bibitem[{Manickathan et~al.(2018)Manickathan, Defraeye, Allegrini, Derome and Carmeliet}]{manickathan2018comparative}
\bibinfo{author}{Manickathan, L.}, \bibinfo{author}{Defraeye, T.}, \bibinfo{author}{Allegrini, J.}, \bibinfo{author}{Derome, D.}, \bibinfo{author}{Carmeliet, J.}, \bibinfo{year}{2018}.
\newblock \bibinfo{title}{Comparative study of flow field and drag coefficient of model and small natural trees in a wind tunnel}.
\newblock \bibinfo{journal}{Urban forestry \& urban greening} \bibinfo{volume}{35}, \bibinfo{pages}{230--239}.
\bibitem[{Massman(1987)}]{massman1987comparative}
\bibinfo{author}{Massman, W.}, \bibinfo{year}{1987}.
\newblock \bibinfo{title}{A comparative study of some mathematical models of the mean wind structure and aerodynamic drag of plant canopies}.
\newblock \bibinfo{journal}{Boundary-layer meteorology} \bibinfo{volume}{40}, \bibinfo{pages}{179--197}.
\bibitem[{Mayhead(1973)}]{mayhead1973some}
\bibinfo{author}{Mayhead, G.}, \bibinfo{year}{1973}.
\newblock \bibinfo{title}{Some drag coefficients for british forest trees derived from wind tunnel studies}.
\newblock \bibinfo{journal}{Agricultural meteorology} \bibinfo{volume}{12}, \bibinfo{pages}{123--130}.
\bibitem[{Mochida et~al.(2008)Mochida, Tabata, Iwata and Yoshino}]{mochida2008examining}
\bibinfo{author}{Mochida, A.}, \bibinfo{author}{Tabata, Y.}, \bibinfo{author}{Iwata, T.}, \bibinfo{author}{Yoshino, H.}, \bibinfo{year}{2008}.
\newblock \bibinfo{title}{Examining tree canopy models for cfd prediction of wind environment at pedestrian level}.
\newblock \bibinfo{journal}{Journal of Wind Engineering and Industrial Aerodynamics} \bibinfo{volume}{96}, \bibinfo{pages}{1667--1677}.
\bibitem[{Molina-Aiz et~al.(2006)Molina-Aiz, Valera, {\'A}lvarez and Madue{\~n}o}]{molina2006wind}
\bibinfo{author}{Molina-Aiz, F.}, \bibinfo{author}{Valera, D.}, \bibinfo{author}{{\'A}lvarez, A.}, \bibinfo{author}{Madue{\~n}o, A.}, \bibinfo{year}{2006}.
\newblock \bibinfo{title}{A wind tunnel study of airflow through horticultural crops: determination of the drag coefficient}.
\newblock \bibinfo{journal}{Biosystems engineering} \bibinfo{volume}{93}, \bibinfo{pages}{447--457}.
\bibitem[{Moradpour et~al.(2017)Moradpour, Afshin and Farhanieh}]{moradpour2017numerical}
\bibinfo{author}{Moradpour, M.}, \bibinfo{author}{Afshin, H.}, \bibinfo{author}{Farhanieh, B.}, \bibinfo{year}{2017}.
\newblock \bibinfo{title}{A numerical investigation of reactive air pollutant dispersion in urban street canyons with tree planting}.
\newblock \bibinfo{journal}{Atmospheric Pollution Research} \bibinfo{volume}{8}, \bibinfo{pages}{253--266}.
\bibitem[{Nebenf{\"u}hr and Davidson(2015)}]{nebenfuhr2015large}
\bibinfo{author}{Nebenf{\"u}hr, B.}, \bibinfo{author}{Davidson, L.}, \bibinfo{year}{2015}.
\newblock \bibinfo{title}{Large-eddy simulation study of thermally stratified canopy flow}.
\newblock \bibinfo{journal}{Boundary-layer meteorology} \bibinfo{volume}{156}, \bibinfo{pages}{253--276}.
\bibitem[{Oke(1989)}]{oke1989micrometeorology}
\bibinfo{author}{Oke, T.R.}, \bibinfo{year}{1989}.
\newblock \bibinfo{title}{The micrometeorology of the urban forest}.
\newblock \bibinfo{journal}{Philosophical Transactions of the Royal Society of London. B, Biological Sciences} \bibinfo{volume}{324}, \bibinfo{pages}{335--349}.
\bibitem[{Oke et~al.(2017)Oke, Mills, Christen and Voogt}]{oke2017urban}
\bibinfo{author}{Oke, T.R.}, \bibinfo{author}{Mills, G.}, \bibinfo{author}{Christen, A.}, \bibinfo{author}{Voogt, J.A.}, \bibinfo{year}{2017}.
\newblock \bibinfo{title}{Urban climates}.
\newblock \bibinfo{publisher}{Cambridge university press}.
\bibitem[{Owens et~al.(2024)Owens, Majumdar, Wilson, Bartholomew and van Reeuwijk}]{owens2024conservative}
\bibinfo{author}{Owens, S.O.}, \bibinfo{author}{Majumdar, D.}, \bibinfo{author}{Wilson, C.E.}, \bibinfo{author}{Bartholomew, P.}, \bibinfo{author}{van Reeuwijk, M.}, \bibinfo{year}{2024}.
\newblock \bibinfo{title}{A conservative immersed boundary method for the multi-physics urban large-eddy simulation model udales v2. 0}.
\newblock \bibinfo{journal}{EGUsphere} \bibinfo{volume}{2024}, \bibinfo{pages}{1--33}.
\bibitem[{Raupach et~al.(1996)Raupach, Finnigan and Brunet}]{raupach1996coherent}
\bibinfo{author}{Raupach, M.R.}, \bibinfo{author}{Finnigan, J.J.}, \bibinfo{author}{Brunet, Y.}, \bibinfo{year}{1996}.
\newblock \bibinfo{title}{Coherent eddies and turbulence in vegetation canopies: the mixing-layer analogy}.
\newblock \bibinfo{journal}{Boundary-Layer Meteorology 25th Anniversary Volume, 1970--1995: Invited Reviews and Selected Contributions to Recognise Ted Munn’s Contribution as Editor over the Past 25 Years} , \bibinfo{pages}{351--382}.
\bibitem[{Ricci et~al.(2022)Ricci, Guasco, Caboni, Orlanno, Giachetta and Repetto}]{ricci2022impact}
\bibinfo{author}{Ricci, A.}, \bibinfo{author}{Guasco, M.}, \bibinfo{author}{Caboni, F.}, \bibinfo{author}{Orlanno, M.}, \bibinfo{author}{Giachetta, A.}, \bibinfo{author}{Repetto, M.}, \bibinfo{year}{2022}.
\newblock \bibinfo{title}{Impact of surrounding environments and vegetation on wind comfort assessment of a new tower with vertical green park}.
\newblock \bibinfo{journal}{Building and Environment} \bibinfo{volume}{207}, \bibinfo{pages}{108409}.
\bibitem[{Rudnicki et~al.(2004)Rudnicki, Mitchell and Novak}]{rudnicki2004wind}
\bibinfo{author}{Rudnicki, M.}, \bibinfo{author}{Mitchell, S.J.}, \bibinfo{author}{Novak, M.D.}, \bibinfo{year}{2004}.
\newblock \bibinfo{title}{Wind tunnel measurements of crown streamlining and drag relationships for three conifer species}.
\newblock \bibinfo{journal}{Canadian Journal of Forest Research} \bibinfo{volume}{34}, \bibinfo{pages}{666--676}.
\bibitem[{Salim et~al.(2015)Salim, Schl{\"u}nzen and Grawe}]{salim2015including}
\bibinfo{author}{Salim, M.H.}, \bibinfo{author}{Schl{\"u}nzen, K.H.}, \bibinfo{author}{Grawe, D.}, \bibinfo{year}{2015}.
\newblock \bibinfo{title}{Including trees in the numerical simulations of the wind flow in urban areas: Should we care?}
\newblock \bibinfo{journal}{Journal of Wind Engineering and Industrial Aerodynamics} \bibinfo{volume}{144}, \bibinfo{pages}{84--95}.
\bibitem[{Santiago et~al.(2017)Santiago, Martilli and Martin}]{santiago2017dry}
\bibinfo{author}{Santiago, J.L.}, \bibinfo{author}{Martilli, A.}, \bibinfo{author}{Martin, F.}, \bibinfo{year}{2017}.
\newblock \bibinfo{title}{On dry deposition modelling of atmospheric pollutants on vegetation at the microscale: Application to the impact of street vegetation on air quality}.
\newblock \bibinfo{journal}{Boundary-layer meteorology} \bibinfo{volume}{162}, \bibinfo{pages}{451--474}.
\bibitem[{Sanz(2003)}]{sanz2003note}
\bibinfo{author}{Sanz, C.}, \bibinfo{year}{2003}.
\newblock \bibinfo{title}{A note on k-$\varepsilon$ modelling of vegetation canopy air-flows}.
\newblock \bibinfo{journal}{Boundary-Layer Meteorology} \bibinfo{volume}{108}, \bibinfo{pages}{191--197}.
\bibitem[{Shaw et~al.(1974)Shaw, Den~Hartog, King and Thurtell}]{shaw1974measurements}
\bibinfo{author}{Shaw, R.}, \bibinfo{author}{Den~Hartog, G.}, \bibinfo{author}{King, K.}, \bibinfo{author}{Thurtell, G.}, \bibinfo{year}{1974}.
\newblock \bibinfo{title}{Measurements of mean wind flow and three-dimensional turbulence intensity within a mature corn canopy}.
\newblock \bibinfo{journal}{Agricultural Meteorology} \bibinfo{volume}{13}, \bibinfo{pages}{419--425}.
\bibitem[{Shaw and Schumann(1992)}]{shaw1992large}
\bibinfo{author}{Shaw, R.H.}, \bibinfo{author}{Schumann, U.}, \bibinfo{year}{1992}.
\newblock \bibinfo{title}{Large-eddy simulation of turbulent flow above and within a forest}.
\newblock \bibinfo{journal}{Boundary-Layer Meteorology} \bibinfo{volume}{61}, \bibinfo{pages}{47--64}.
\bibitem[{Shu et~al.(2020)Shu, Wang and Mortezazadeh}]{shu2020dimensional}
\bibinfo{author}{Shu, C.}, \bibinfo{author}{Wang, L.L.}, \bibinfo{author}{Mortezazadeh, M.}, \bibinfo{year}{2020}.
\newblock \bibinfo{title}{Dimensional analysis of reynolds independence and regional critical reynolds numbers for urban aerodynamics}.
\newblock \bibinfo{journal}{Journal of Wind Engineering and Industrial Aerodynamics} \bibinfo{volume}{203}, \bibinfo{pages}{104232}.
\bibitem[{Smith et~al.(2021)Smith, Bentrup, Kellerman, MacFarland, Straight and Ameyaw}]{smith2021windbreaks}
\bibinfo{author}{Smith, M.M.}, \bibinfo{author}{Bentrup, G.}, \bibinfo{author}{Kellerman, T.}, \bibinfo{author}{MacFarland, K.}, \bibinfo{author}{Straight, R.}, \bibinfo{author}{Ameyaw, L.}, \bibinfo{year}{2021}.
\newblock \bibinfo{title}{Windbreaks in the united states: A systematic review of producer-reported benefits, challenges, management activities and drivers of adoption}.
\newblock \bibinfo{journal}{Agricultural Systems} \bibinfo{volume}{187}, \bibinfo{pages}{103032}.
\bibitem[{Suter et~al.(2022)Suter, Grylls, S{\"u}tzl, Owens, Wilson and van Reeuwijk}]{suter2022udales}
\bibinfo{author}{Suter, I.}, \bibinfo{author}{Grylls, T.}, \bibinfo{author}{S{\"u}tzl, B.S.}, \bibinfo{author}{Owens, S.O.}, \bibinfo{author}{Wilson, C.E.}, \bibinfo{author}{van Reeuwijk, M.}, \bibinfo{year}{2022}.
\newblock \bibinfo{title}{udales 1.0: a large-eddy simulation model for urban environments}.
\newblock \bibinfo{journal}{Geoscientific Model Development} \bibinfo{volume}{15}, \bibinfo{pages}{5309--5335}.
\bibitem[{Svensson and H{\"a}ggkvist(1990)}]{svensson1990two}
\bibinfo{author}{Svensson, U.}, \bibinfo{author}{H{\"a}ggkvist, K.}, \bibinfo{year}{1990}.
\newblock \bibinfo{title}{A two-equation turbulence model for canopy flows}.
\newblock \bibinfo{journal}{Journal of wind engineering and industrial aerodynamics} \bibinfo{volume}{35}, \bibinfo{pages}{201--211}.
\bibitem[{Vollsinger et~al.(2005)Vollsinger, Mitchell, Byrne, Novak and Rudnicki}]{vollsinger2005wind}
\bibinfo{author}{Vollsinger, S.}, \bibinfo{author}{Mitchell, S.J.}, \bibinfo{author}{Byrne, K.E.}, \bibinfo{author}{Novak, M.D.}, \bibinfo{author}{Rudnicki, M.}, \bibinfo{year}{2005}.
\newblock \bibinfo{title}{Wind tunnel measurements of crown streamlining and drag relationships for several hardwood species}.
\newblock \bibinfo{journal}{Canadian Journal of Forest Research} \bibinfo{volume}{35}, \bibinfo{pages}{1238--1249}.
\bibitem[{Vranckx et~al.(2015)Vranckx, Vos, Maiheu and Janssen}]{vranckx2015impact}
\bibinfo{author}{Vranckx, S.}, \bibinfo{author}{Vos, P.}, \bibinfo{author}{Maiheu, B.}, \bibinfo{author}{Janssen, S.}, \bibinfo{year}{2015}.
\newblock \bibinfo{title}{Impact of trees on pollutant dispersion in street canyons: A numerical study of the annual average effects in antwerp, belgium}.
\newblock \bibinfo{journal}{Science of the Total Environment} \bibinfo{volume}{532}, \bibinfo{pages}{474--483}.
\bibitem[{Vreman(2004)}]{vreman2004eddy}
\bibinfo{author}{Vreman, A.}, \bibinfo{year}{2004}.
\newblock \bibinfo{title}{An eddy-viscosity subgrid-scale model for turbulent shear flow: Algebraic theory and applications}.
\newblock \bibinfo{journal}{Physics of fluids} \bibinfo{volume}{16}, \bibinfo{pages}{3670--3681}.
\bibitem[{Wang et~al.(1996)Wang, Plate, Rau and Keiser}]{wang1996scale}
\bibinfo{author}{Wang, Z.Y.}, \bibinfo{author}{Plate, E.J.}, \bibinfo{author}{Rau, M.}, \bibinfo{author}{Keiser, R.}, \bibinfo{year}{1996}.
\newblock \bibinfo{title}{Scale effects in wind tunnel modelling}.
\newblock \bibinfo{journal}{Journal of Wind Engineering and Industrial Aerodynamics} \bibinfo{volume}{61}, \bibinfo{pages}{113--130}.
\bibitem[{Weninger et~al.(2021)Weninger, Scheper, Lack{\'o}ov{\'a}, Kitzler, Gartner, King, Cornelis, Strauss and Michel}]{weninger2021ecosystem}
\bibinfo{author}{Weninger, T.}, \bibinfo{author}{Scheper, S.}, \bibinfo{author}{Lack{\'o}ov{\'a}, L.}, \bibinfo{author}{Kitzler, B.}, \bibinfo{author}{Gartner, K.}, \bibinfo{author}{King, N.}, \bibinfo{author}{Cornelis, W.}, \bibinfo{author}{Strauss, P.}, \bibinfo{author}{Michel, K.}, \bibinfo{year}{2021}.
\newblock \bibinfo{title}{Ecosystem services of tree windbreaks in rural landscapes—a systematic review}.
\newblock \bibinfo{journal}{Environmental Research Letters} \bibinfo{volume}{16}, \bibinfo{pages}{103002}.
\bibitem[{Wilson et~al.(1982)Wilson, Ward, Thurtell and Kidd}]{wilson1982statistics}
\bibinfo{author}{Wilson, J.}, \bibinfo{author}{Ward, D.}, \bibinfo{author}{Thurtell, G.}, \bibinfo{author}{Kidd, G.}, \bibinfo{year}{1982}.
\newblock \bibinfo{title}{Statistics of atmospheric turbulence within and above a corn canopy}.
\newblock \bibinfo{journal}{Boundary-Layer Meteorology} \bibinfo{volume}{24}, \bibinfo{pages}{495--519}.
\bibitem[{Wilson(1985)}]{wilson1985numerical}
\bibinfo{author}{Wilson, J.D.}, \bibinfo{year}{1985}.
\newblock \bibinfo{title}{Numerical studies of flow through a windbreak}.
\newblock \bibinfo{journal}{Journal of Wind Engineering and Industrial Aerodynamics} \bibinfo{volume}{21}, \bibinfo{pages}{119--154}.
\bibitem[{Yang et~al.(2017)Yang, Juan, Wen and Chang}]{yang2017numerical}
\bibinfo{author}{Yang, A.S.}, \bibinfo{author}{Juan, Y.H.}, \bibinfo{author}{Wen, C.Y.}, \bibinfo{author}{Chang, C.J.}, \bibinfo{year}{2017}.
\newblock \bibinfo{title}{Numerical simulation of cooling effect of vegetation enhancement in a subtropical urban park}.
\newblock \bibinfo{journal}{Applied energy} \bibinfo{volume}{192}, \bibinfo{pages}{178--200}.

\end{thebibliography}

\end{document}